\documentclass[12pt,preprint]{aastex}

\newcommand{\teff}{$T_{\rm eff}$} 
\newcommand{\kms}{km s$^{-1}$}
\begin{document}

\title{Heavy element abundances in giant stars of the globular 
clusters M4 and M5\altaffilmark{1}}

\author{David Yong}
\affil{Research School of Astronomy and Astrophysics, Australian National 
University, Mount Stromlo Observatory, Cotter Road, Weston Creek, ACT 2611, 
Australia}
\email{yong@mso.anu.edu.au}

\author{Amanda I.\ Karakas}
\affil{Research School of Astronomy and Astrophysics, Australian National 
University, Mount Stromlo Observatory, Cotter Road, Weston Creek, ACT 2611, 
Australia}
\email{akarakas@mso.anu.edu.au}

\author{David L.\ Lambert}
\affil{The W.\ J.\ McDonald Observatory, University of Texas, Austin, TX 78712}
\email{dll@astro.as.utexas.edu}

\author{Alessandro Chieffi}
\affil{Istituto Nazionale di Astrofisica - Istituto di Astrofisica 
Spaziale e Fisica Cosmica, Via Fosso del Cavaliere, I-00133, Roma, Italy}
\email{alessandro.chieffi@iasf-roma.inaf.it}

\author{Marco Limongi}
\affil{Istituto Nazionale di Astrofisica - Osservatorio Astronomico 
di Roma, Via Frascati 33, I-00040, Roma, Italy}
\email{marco@oa-roma.inaf.it}

\altaffiltext{1}{Based on 
observations made with the Magellan Clay Telescope at Las Campanas 
Observatory.}

\begin{abstract}

We present a comprehensive abundance analysis of 27 heavy elements in 
bright giant stars of the globular clusters M4 and M5 based on high resolution, 
high signal-to-noise ratio  
spectra obtained with the Magellan Clay Telescope. 
We confirm and expand upon previous results for these clusters by 
showing that (1) all elements heavier than, and including, 
Si have constant abundances within each 
cluster, (2) the elements from Ca to Ni have  
indistinguishable compositions in M4 and M5, 
(3) Si, Cu, Zn, and  all $s$-process 
elements are approximately 0.3 dex overabundant in M4 relative to M5, and (4) 
the $r$-process elements Sm, Eu, Gd, and Th are slightly overabundant 
in M5 relative to M4. 
The cluster-to-cluster abundance differences for Cu and Zn are 
intriguing, especially in light of their uncertain nucleosynthetic origins. 
We confirm that stars other than Type Ia supernovae must produce significant 
amounts of Cu and Zn at or below the clusters' metallicities. 
If intermediate-mass AGB stars or massive stars are responsible for the 
Cu and Zn enhancements in M4, the similar 
[Rb/Zr] ratios and (preliminary) Mg isotope ratios in both clusters 
may be problematic for either scenario. 
For the elements from Ba to Hf, we assume that the $s$- and $r$-process 
contributions are scaled versions of the solar $s$- and $r$-process 
abundances. We quantify the relative fractions 
of $s$- and $r$-process material 
for each cluster and show that 
they provide an excellent fit to the observed abundances. 

\end{abstract}

\keywords{Galaxy: Abundances, Galaxy: Globular Clusters: Individual: Messier 
Number: M4, Galaxy: Globular Clusters: Individual: Messier Number: M5, 
Stars: Abundances}

\section{Introduction}
\label{sec:intro}

Globular clusters continue to play a vital role in testing many aspects 
of stellar evolution and stellar nucleosynthesis. Observational 
and theoretical studies of globular clusters 
have focused heavily upon (1) the star-to-star 
light element abundance variations 
\citep{cottrell81,langer95,gratton04}, 
(2) the cluster-to-cluster variation in the 
color distribution of horizontal branch stars, the so-called 
``2nd parameter effect'' 
\citep{sandage67,lee94,carretta06}, 
and (3) the multiple populations as inferred from large spreads 
in metallicity and/or detailed structure in color-magnitude diagrams 
\citep{butler78,norris95b,bekki06}. 

Abundance measurements of the 
$s$-process and $r$-process elements offer great insight into 
stellar nucleosynthesis and globular cluster chemical evolution. 
Aside from M15, a metal-poor cluster that displays a 
scaled-solar $r$-process abundance 
distribution \citep{sneden97,sneden00b,otsuki06}, 
in general only 
a handful of $s$-process elements (e.g., Y, Zr, Ba, La) and the $r$-process 
element Eu have been measured in globular clusters.  

The globular clusters M4 and M5 are particularly well suited for 
refining our understanding of stellar evolution and stellar 
nucleosynthesis, especially for the neutron-capture elements. 
\citet{M4,M5} showed that these clusters have essentially
identical metallicities, [Fe/H] = $-$1.2, based on high resolution 
spectra of large samples, 36 stars in each cluster. 
They also showed that the abundance similarities for these clusters 
extend to numerous $\alpha$- and Fe-peak elements as well as the $r$-process 
element Eu. However, 
the $s$-process elements revealed striking abundance differences between 
these two clusters. Specifically, the heavy $s$-process elements Ba and La are 
overabundant in M4 relative to M5 \citep{M4,M5}. With the notable 
exception of $\omega$ Cen, M4 may be uniquely 
enriched in $s$-process elements among the Galactic globular 
clusters \citep{pritzl05}. 

\citet{rbpbm4m5} recently extended the analysis of neutron-capture elements in 
these two clusters to Rb and Pb, $s$-process 
elements which may be overproduced in metal-poor 
asymptotic giant branch (AGB) stars \citep{busso99,travaglio01}. 
Whereas Pb production is dominated by 2 to 4$M_{\odot}$ low-metallicity 
AGB stars \citep{travaglio01}, 
the intermediate-mass 4 to 8$M_{\odot}$ AGB stars are predicted 
to dominate Rb production \citep{vanraai08}. 
Not 
surprisingly, M4 again had higher abundance ratios [Rb/Fe] and [Pb/Fe] 
than M5. However, the abundance ratios [Rb/X] for X = Y, Zr, and La were very 
similar in the two clusters indicating that the nature of the $s$-process 
products is very similar for both clusters but that M4 formed from 
gas with a higher concentration of these products. 
A comprehensive study of $s$-process elements in these 
two clusters promises to provide a novel observational study 
of the $s$-process at low metallicities as well as valuable clues 
to the chemical evolution diversity of globular clusters. 
In this paper we present such an analysis, 
focusing upon a suite of $\alpha$-, Fe-peak, $s$-process,
and $r$-process elements. 

\section{Observations and analysis}
\label{sec:data}

The sample consists of 12 bright giants in M4 and 2 bright giants in 
M5 observed with the high resolution spectrograph MIKE \citep{mike} on 
the Magellan Clay Telescope. These are the same high quality spectra 
analyzed by \citet{rbpbm4m5} 
(R = 55,000, wavelength coverage from 3800\AA\ to 8500\AA, and 
S/N = 140 per resolution element at 4000\AA\ and S/N = 800 
per resolution element at 7800\AA). 

The stellar parameters are presented in Table \ref{tab:param} 
and were determined using a traditional 
spectroscopic approach. Equivalent widths (EWs) for a set of Fe 
lines were measured using routines in 
IRAF\footnote{IRAF (Image Reduction and Analysis
Facility) is distributed by the National Optical Astronomy
Observatory, which is operated by the Association of Universities
for Research in Astronomy, Inc., under contract with the National
Science Foundation.}. We used the local thermodynamic equilibrium (LTE) 
stellar line analysis program 
MOOG \citep{moog} and LTE model atmospheres from 
the \citet{kurucz93} grid to derive an abundance for a given line. 
The effective temperature, \teff, was adjusted 
until the abundances from Fe\,{\sc i} lines displayed no trend with 
the lower excitation potential. The surface gravity, $\log g$, was 
adjusted until the abundances from Fe\,{\sc i} and Fe\,{\sc ii} lines 
were in agreement. The microturbulent velocity, $\xi_t$, was adjusted 
until there was no trend between the abundances from Fe\,{\sc i} 
lines and EW. This process was iterated until self consistent 
stellar parameters were obtained. 
Ideally, the trends between abundance and EW and between abundance 
and lower excitation potential 
should be exactly zero. Further, the abundances $\log\epsilon$(Fe\,{\sc i}) 
and $\log\epsilon$(Fe\,{\sc ii}) should be exactly the same. 
In our analysis, we explored stellar parameters at discrete values. 
For \teff, we considered values at every 25 K (e.g., 4000 K, 4025 K, etc), 
for $\log g$, we considered values at every 0.05 dex 
(e.g., 1.00 dex, 1.05 dex, etc), 
and for $\xi_t$, we considered values at every 0.05 \kms\ (e.g., 1.70 \kms, 
1.75 \kms\, etc). 
We assumed that excitation equilibrium was satisfied when the 
slope between $\log\epsilon$(Fe\,{\sc i}) and lower excitation potential 
was $\le 0.004$. We assumed that ionization equilibrium was achieved when 
$| \log\epsilon$(Fe\,{\sc i}) $-$ $\log\epsilon$(Fe\,{\sc ii}) $| \le$ 0.02 
dex. The microturbulent velocity was set when the slope between 
$\log\epsilon$(Fe\,{\sc i}) and reduced equivalent width ($\log W/\lambda$) 
was $\le 0.004$. 
Our stellar parameters are in 
good agreement with \citet{M4,M5}. 
We estimate that the internal 
errors are \teff\ $\pm$ 50 K, $\log g$ $\pm$ 0.2 dex, and 
$\xi_t$ $\pm$ 0.2 \kms. For further 
details regarding the observations, 
data reduction, and derivation of stellar parameters, see 
\citet{rbpbm4m5}. 

\begin{figure}
\epsscale{0.8}
\plotone{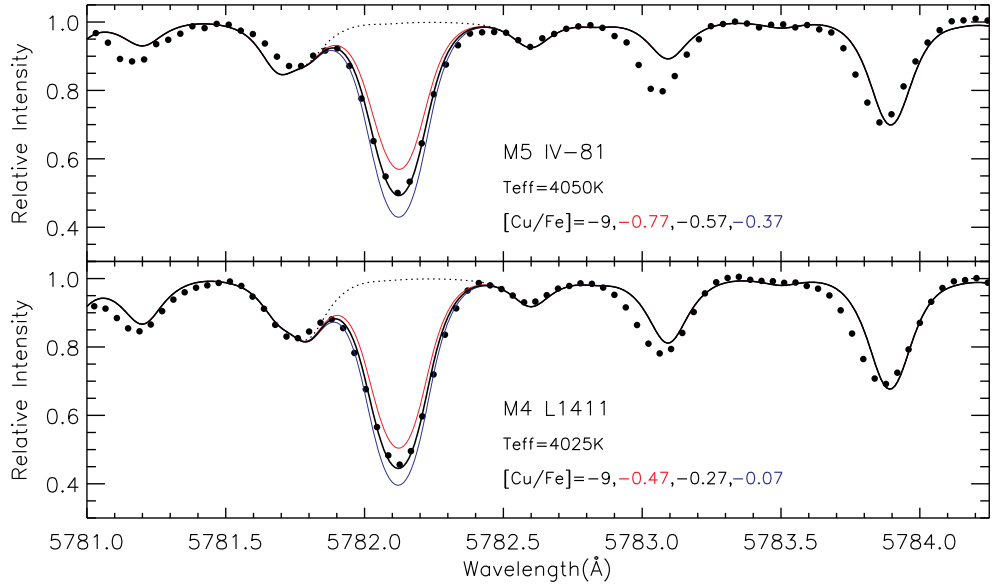}
\caption{Observed spectra (filled circles) and synthetic spectra 
(solid and dotted lines) for M5 IV-81 (upper) and M4 L1411 (lower) near the 
5782\AA\ Cu\,{\sc i} line. The synthetic spectra show the best fit (thick 
black line), unsatisfactory fits $\pm$ 0.2 dex (thin red and blue 
lines), and a fit with no Cu (dotted line). 
\label{fig:cu}}
\end{figure}

\begin{figure}
\epsscale{0.8}
\plotone{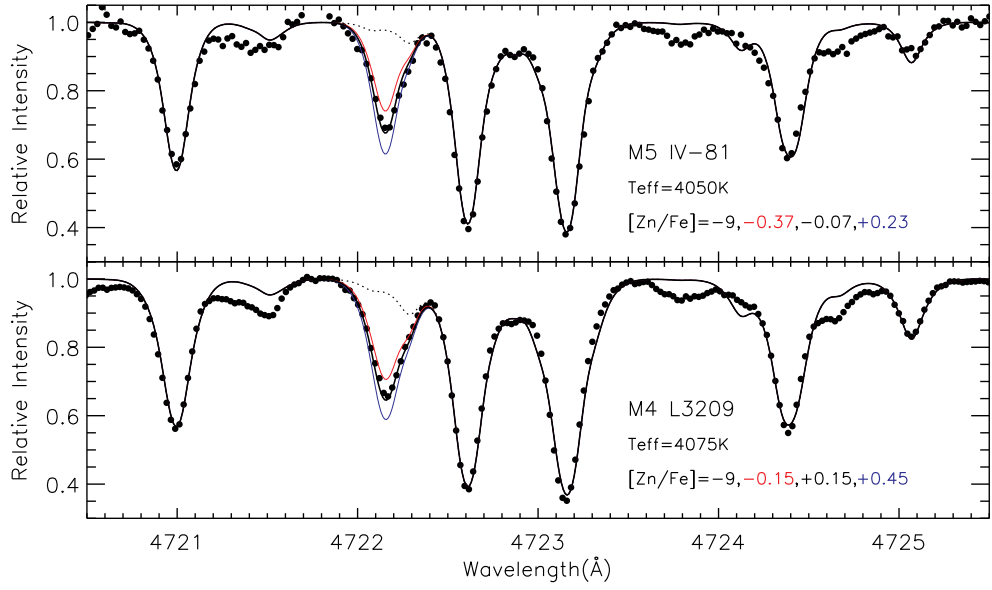}
\caption{Same as Figure \ref{fig:cu} but for the 4722\AA\ Zn\,{\sc i} line 
in stars M5 IV-81 and M4 L3209 
(unsatisfactory fits $\pm$ 0.3 dex are shown). 
\label{fig:zn}}
\end{figure}

\begin{figure}
\epsscale{0.8}
\plotone{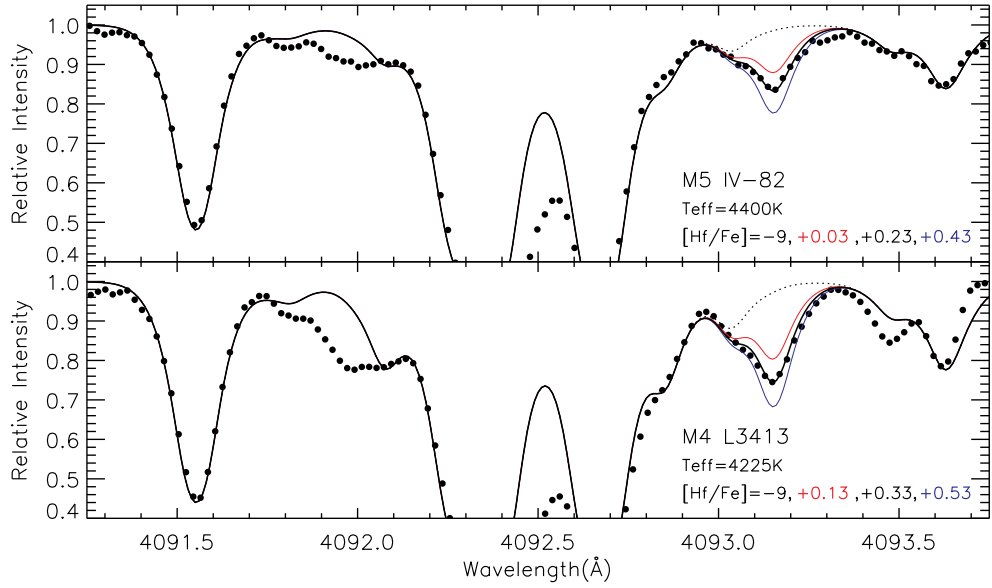}
\caption{Same as Figure \ref{fig:cu} but for the 4093\AA\ Hf\,{\sc ii} 
line in stars M5 IV-82 and M4 L3413 
(unsatisfactory fits $\pm$ 0.2 dex are shown). 
\label{fig:hf}}
\end{figure}

Whenever possible, the abundances for additional elements were determined 
via equivalent width analysis. For particular lines, the 
abundances were determined by generating synthetic spectra using MOOG 
and adjusting the abundance to match the observed spectrum. 
In Figures \ref{fig:cu} to \ref{fig:th}, 
we show examples of the abundance determination 
via synthetic spectra. 
When required, isotopic and/or hyperfine splitting was taken into account. 
For Pr, we were unable to obtain sufficient information 
to appropriately treat the hyperfine 
splitting. The EWs for the 5322.76\AA\ Pr\,{\sc ii} line 
range from 24.6 to 65.4 m\AA\ and we 
caution that the abundances may be slightly overestimated. 
The line list, source of $gf$ values, and EWs are 
given in Table \ref{tab:ew}. The adopted solar 
abundances are given in Table \ref{tab:solar}. 
The final abundances are presented in 
Tables \ref{tab:abund1} and \ref{tab:abund2}. 
The abundances for Rb, Y, Zr, La, Eu, and Pb were taken 
from \citet{rbpbm4m5}. The abundance dependences upon the 
model parameters are given in Table \ref{tab:parvar}. 

For the elements Si, Ca, Sc, Ti, V, Fe, Ni, La, and Eu, 
our mean abundances [X/Fe] and [Fe/H] 
for M4 and for M5 are in good agreement with the 
mean cluster values measured by \citet{M4,M5} (see Figure \ref{fig:comp}). 
For Mn and Cu, our mean abundances [X/Fe] for 
M4 and M5 are also in good agreement with the mean cluster values 
measured by \citet{sobeck06} and \citet{simmerer03} respectively, 
who used the same spectra and stellar 
parameters as \citet{M4,M5}. 
For all elements previously measured in these clusters, the mean abundance 
differences are $\Delta$[A/B]$_{\rm (This~study~-~Literature)}$ = 
0.08 dex ($\sigma$ = 0.08 dex) and 0.09 dex ($\sigma$ = 0.10 dex) 
for M4 and M5 respectively. 
However, when we consider the mean cluster abundance differences, 
([A/B]$^{\rm M4}_{\rm This~study} -$ [A/B]$^{\rm M5}_{\rm This~study}$) $-$
([A/B]$^{\rm M4}_{\rm Literature} -$ [A/B]$^{\rm M5}_{\rm Literature}$), 
our results are in excellent agreement, $-$0.01 dex ($\sigma$ = 0.06 dex), 
with \citet{M4,M5}, \citet{sobeck06}, and \citet{simmerer03}. 
The similarity in the abundance differences, M4 $-$ M5, between the various 
studies highlights the differential nature of the analyses. That is, the 
various studies have derived abundances for stars in M4 and M5, that 
cover a small range of stellar parameters, using homogeneous 
spectra and analysis techniques. While the similarity in abundance 
differences is very pleasing, it is not unexpected. We will take 
advantage of the very precise abundance differences to explore the 
chemical similarities and differences between these two clusters. 

\section{Results}
\label{sec:results}

In Figures \ref{fig:ab1} to \ref{fig:ab3}, the upper panels 
show the mean abundances [X/Fe] and [Fe/H] for 
M4 and M5 and the lower panels show the abundance differences, 
$\Delta$[A/B] = [A/B]$_{\rm M4}$ $-$ [A/B]$_{\rm M5}$. 
The error bars in these figures represent the observed spread ($\sigma$) 
in the measured abundances. We note that the 
observed spread in the measured abundances is in excellent agreement 
with the predicted spread, 
$\sigma_{\rm predicted} - \sigma_{\rm observed~(M4)} = 0.00 $ 
($\sigma$ = 0.05) and 
$\sigma_{\rm predicted} - \sigma_{\rm observed~(M5)} = 0.03 $ 
($\sigma$ = 0.05). For all elements in this study, we therefore 
regard the abundances as being 
constant within each cluster. 
(Note that the $\sigma_{\rm predicted}$, taken directly from 
Table \ref{tab:parvar}, neglect errors due to EW measurements and 
continuum placement which may be small in our high quality spectra 
of moderately metal-poor stars. Further, these uncertainties only 
account for the 
relative internal uncertainties.) For M4, the light element Na was found 
to vary by $\Delta$[Na/Fe] = 0.62 dex within our sample \citep{rbpbm4m5}. 
In this study, we find that 
none of the abundance ratios [X/Fe] for X = Si to Th are 
correlated with [Na/Fe]. 

\subsection{The $\alpha$- and Fe-peak elements, Si to Zn}

In Figure \ref{fig:ab1}, we plot the abundance ratios [X/Fe] and 
[Fe/H] for various $\alpha$- and Fe-peak elements from Si to Zn. 
We confirm the findings by \citet{M4,M5} that M4 and M5 have very similar 
abundances of Ca, Sc, Ti, V, Fe, and Ni. 
Our results extend the abundance similarities to the elements 
Cr and Co. For every element between Ca and Ni inclusive, M4 and M5 
have essentially identical compositions, 
$<$$\Delta$[A/B]$_{\rm (M4-M5)}$$>$ = 0.04 ($\sigma$ = 0.06). 
The remarkable abundance similarity between M4 and M5 
for the elements from Ca to Ni is most 
readily seen in the lower panel of Figure \ref{fig:ab1}. 
In Figures \ref{fig:speccomp1} to \ref{fig:speccomp3}, we plot 
the spectra for M4 L1411 and M5 IV-81, two stars for which we 
derived very similar values of \teff. Therefore, any difference 
between the spectra is most likely due to abundance differences 
between the two stars. In these figures, the similarities in the line 
strengths of Ca, Sc, Ti, V, Fe, 
and Co reinforce our findings that these abundances (and 
the stellar parameters) are very similar in the two stars. 

The abundance similarities for M4 and M5 
do not extend to the elements Si, Cu, and 
Zn. For these elements, M4 has [X/Fe] ratios roughly 0.3 dex higher than 
those in M5. 

The case of Si was already documented by \citet{M5}. 
In Figure \ref{fig:speccomp1}, lines of Si show considerably 
different strengths in two stars with very similar values of \teff, 
which supports the claim that the Si abundances differ 
between M4 L1411 and M5 IV-81. Curiously, the behavior of 
Si is very different from the other $\alpha$-elements 
Ca and Ti. 

The differences in Zn abundances between M4 L3209 and M5 IV-81 
can be seen in Figure \ref{fig:zn}. 
Although the abundance differences are not immediately obvious when 
visually comparing 
line strengths in these two stars, 
synthetic spectra reveal that M4 L3209 is overabundant in 
Zn relative to M5 IV-81. 

For the abundances of Cu, differences between M4 and M5 were 
reported by \citet{simmerer03}. In Figure \ref{fig:cu}, the 
abundance differences between M4 L1411 and M5 IV-81 
can be seen from the synthetic spectra. 
\citet{simmerer03} 
suggested that the [Cu/Fe] 
abundances in both clusters were in agreement with field stars at the 
same metallicity. (They found a 0.17 dex difference in [Fe/H] between these 
clusters with M4 being the more metal-rich cluster). 
Inspection of their Figure 6 shows that while 
[Cu/Fe] in halo field stars (drawn exclusively from 
\citealt{mishenina02}) appears to be falling rapidly with decreasing 
metallicity, there are no field stars in the metallicity regime between 
M4 and M5. The sparse data set reveal that 
for the five field halo stars 
closest in metallicity to M4 and M5, 
the [Cu/Fe] values cover roughly 0.6 dex. 
However, some of the \citet{mishenina02} comparison field halo 
stars may be regarded as being (slightly) chemically 
peculiar. 
HD 6833 has a higher than usual ratio [Zr/Fe] = +0.48 \citep{fulbright00}
and HD 166161 has a higher than usual ratio 
$\log\epsilon$(La/Eu) = 0.71 \citep{simmerer04}. 
For these two stars, the [Cu/Fe] ratios may not be representative of 
the general halo trend and therefore, the claim that M4 and M5 
have Cu abundances in agreement with field halo stars needs to be
re-examined. 
Recent Cu measurements by \citet{primas08} 
more clearly define the [Cu/Fe] versus [Fe/H] trend in field halo stars. 
Their results 
suggest that the 0.25 dex difference in [Cu/Fe] between M4 and M5 may 
exceed the intrinsic spread for the small range in 
[Fe/H] spanned by the two clusters. However, we note that such 
abundance comparisons between different studies can be problematic 
due to (unknown) systematic differences.  

\subsection{The $s$-process and $r$-process elements, Rb to Th}

In Figure \ref{fig:ab2}, we plot the abundance ratios [X/Fe] for 
various $s$-process and $r$-process elements from Rb to Gd and 
in Figure \ref{fig:ab3}, we plot the abundance ratios for all elements 
including Hf, Pb, and Th. 
\citet{M4,M5} showed that the $s$-process elements Ba and La are 
overabundant in M4 relative to M5 and \citet{rbpbm4m5} extended 
the abundance differences to the $s$-process elements Rb and Pb. 
We find that every $s$-process element 
shows overabundances 
in M4 relative to M5. In Figures \ref{fig:speccomp1} (Zr), 
\ref{fig:speccomp2} (Y and Mo), and \ref{fig:speccomp3} 
(Y, Zr, and Nd), various $s$-process elements have considerably 
different line strengths reinforcing the claim that 
the abundances differ between 
M4 L1411 and M5 IV-81. The nine $s$-process elements 
measured in M4 and M5, Rb, Sr, Y, Zr, Mo, Ba, La, Ce, and Pb, 
have $<$$\Delta$[X/Fe]$_{\rm (M4-M5)}$$>$ = 0.38 ($\sigma$ = 0.14). 
Although the abundances are derived from a single line in a crowded 
region, the behavior of Pb, $\Delta$[Pb/Fe]$_{\rm (M4-M5)}$ = 0.65, may differ 
from that of the other $s$-elements in these clusters. 
Recent observations of Pb by \citet{aoki08} in a 
sample of field halo stars without large 
carbon enhancements show that M5 has a [Pb/Fe] ratio comparable to 
field stars at the same metallicity. 
The \citet{aoki08} results also show that the globular clusters 
NGC 6752 and M13 have [Pb/Fe] ratios typical of field halo stars 
at the same metallicity \citep{rbpbsubaru}. 

\begin{figure}
\epsscale{0.8}
\plotone{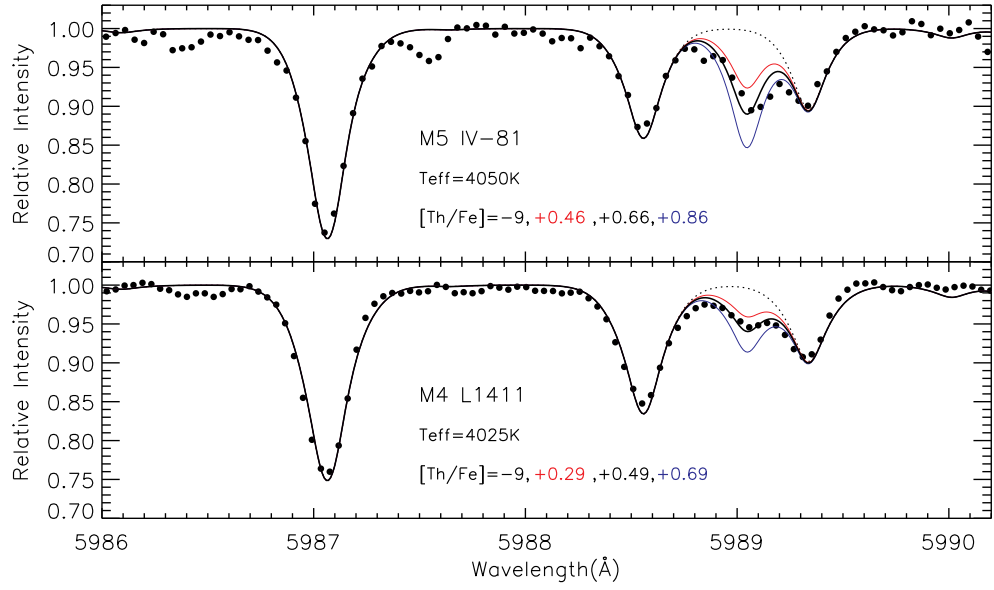}
\caption{Same as Figure \ref{fig:cu} but for the 5989\AA\ Th\,{\sc ii}
line (unsatisfactory fits $\pm$ 0.2 dex are shown). 
\label{fig:th}}
\end{figure}

\begin{figure}
\epsscale{0.8}
\plotone{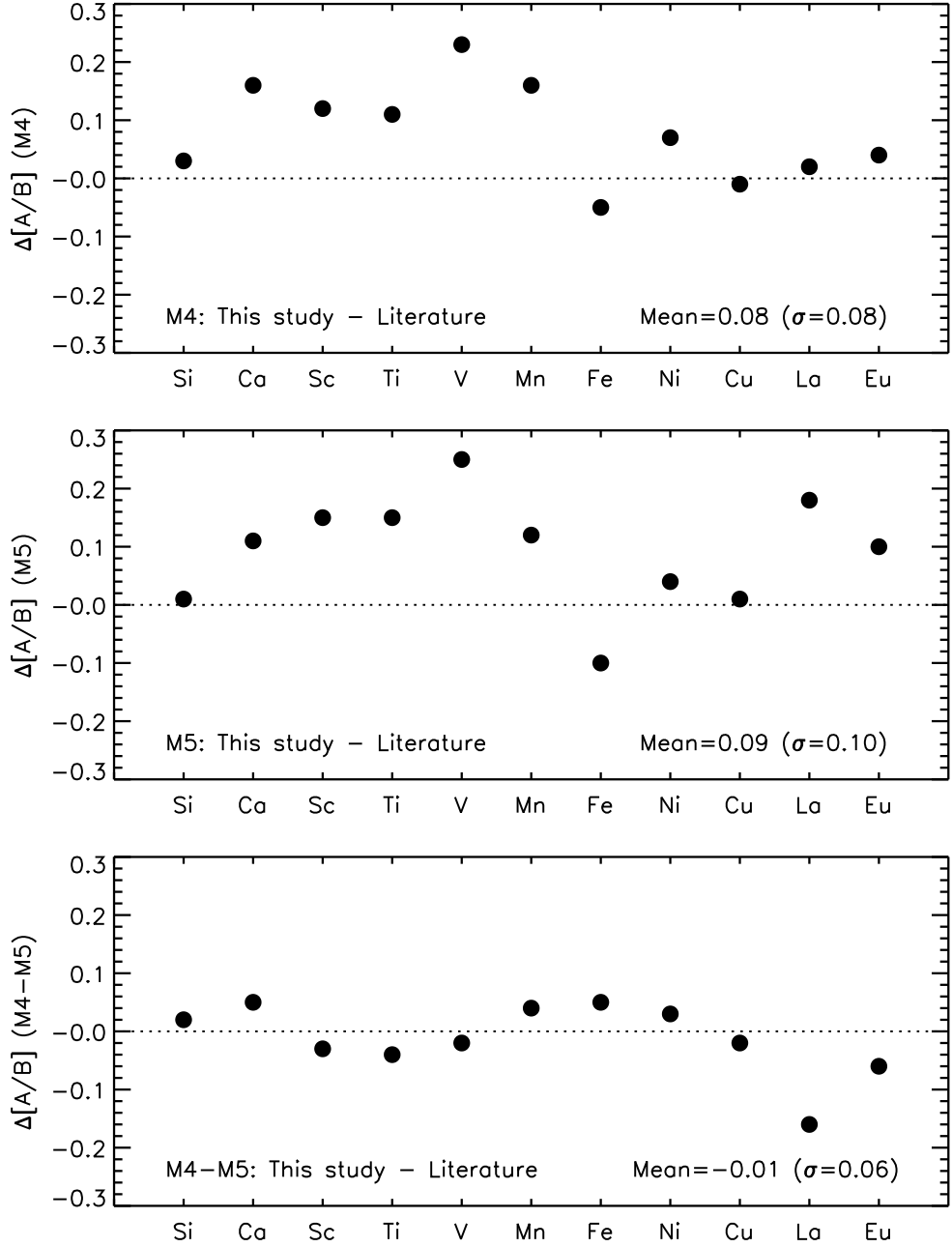}
\caption{A comparison of the mean cluster values [X/Fe] and [Fe/H] 
between this study 
and the literature for M4 (upper), M5 (middle), and M4$-$M5 (lower). 
Literature values are from \citet{M4,M5} except Mn \citep{sobeck06} and 
Cu \citep{simmerer03}. The mean differences and standard deviations 
are shown. 
\label{fig:comp}}
\end{figure}

\begin{figure}
\epsscale{0.8}
\plotone{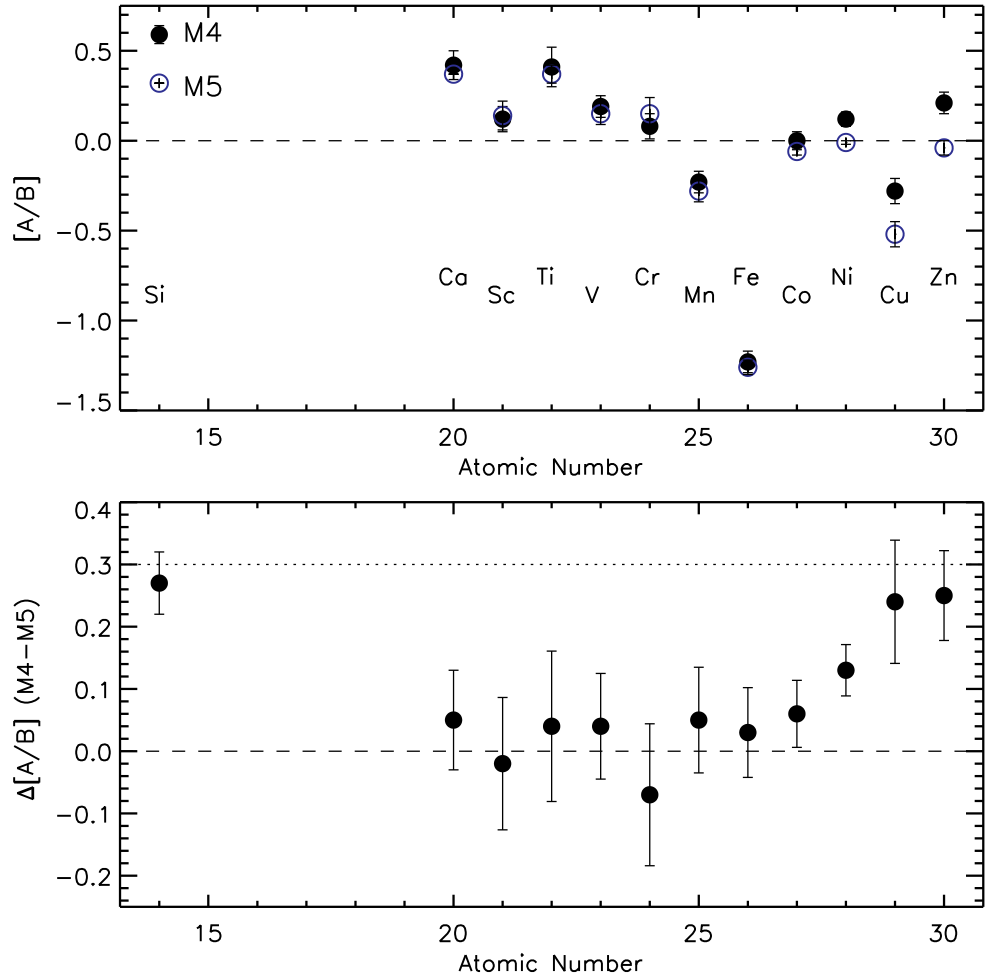}
\caption{The upper panel shows [X/Fe] for the elements Si to Zn as well 
as [Fe/H] for M4 (filled circles) and M5 (open blue circles). 
The lower panel shows the difference in abundance ratios, 
$\Delta$[X/Fe] (M4$-$M5). \label{fig:ab1}}
\end{figure}

\begin{figure}
\epsscale{0.8}
\plotone{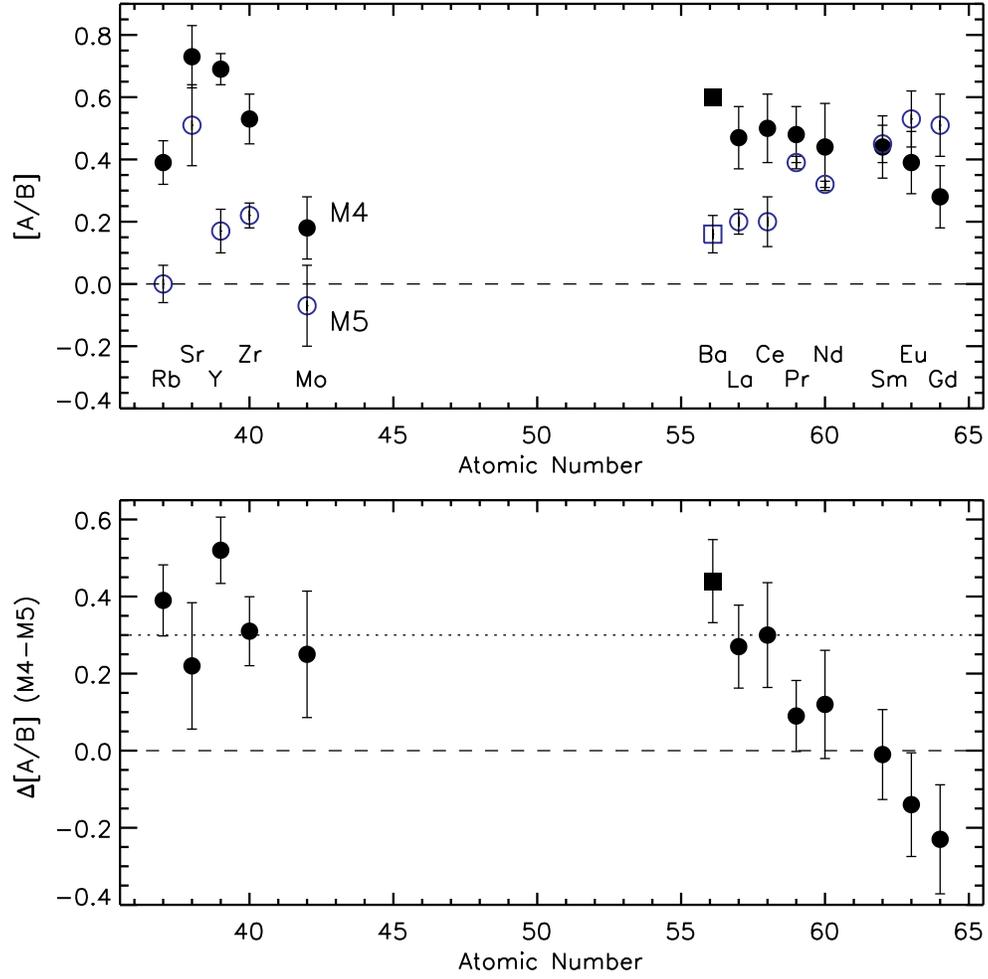}
\caption{The same as Figure \ref{fig:ab1} but for the 
elements Rb to Gd. All data points are from this study and 
\citet{rbpbm4m5}, except 
Ba which is taken from \citet{M4,M5}. \label{fig:ab2}}
\end{figure}

\begin{figure}
\epsscale{0.8}
\plotone{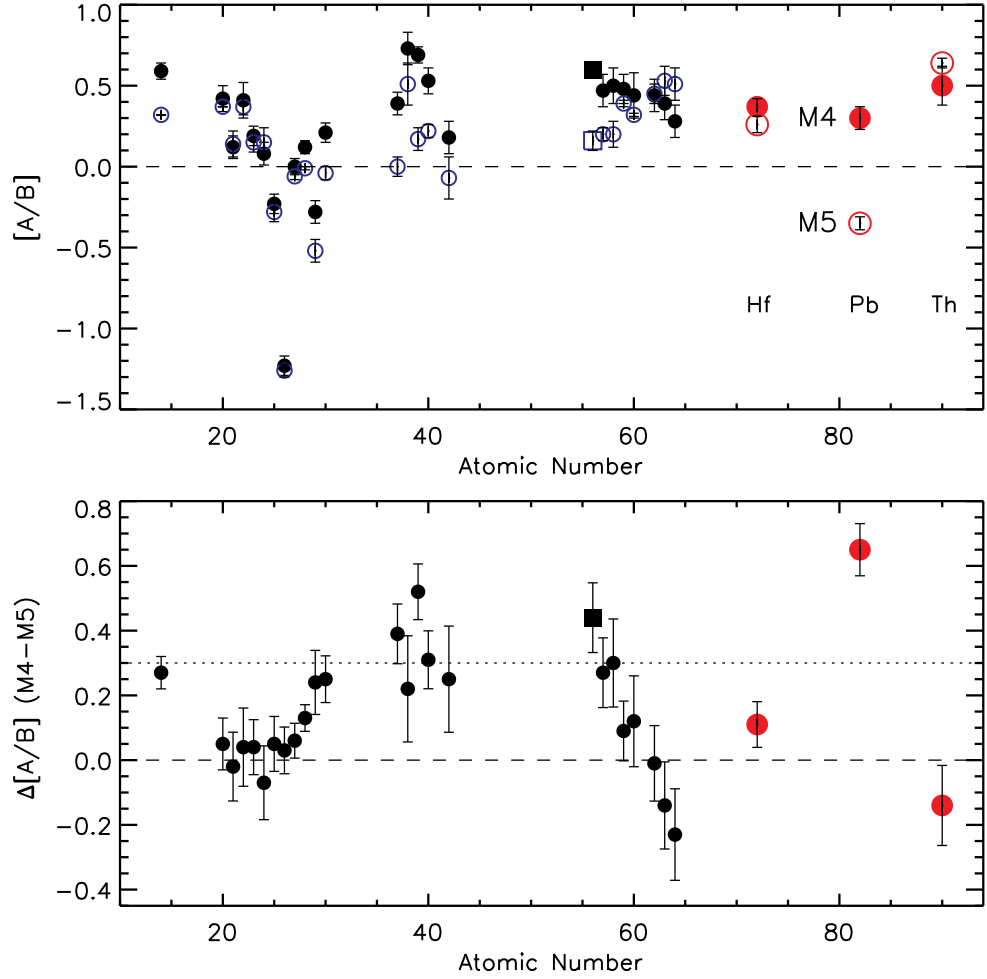}
\caption{The same as Figures \ref{fig:ab1} and \ref{fig:ab2} but for all 
elements. The heavy neutron-capture elements Hf, Pb, and Th 
are highlighted in red. \label{fig:ab3}}
\end{figure}

\begin{figure}
\epsscale{0.8}
\plotone{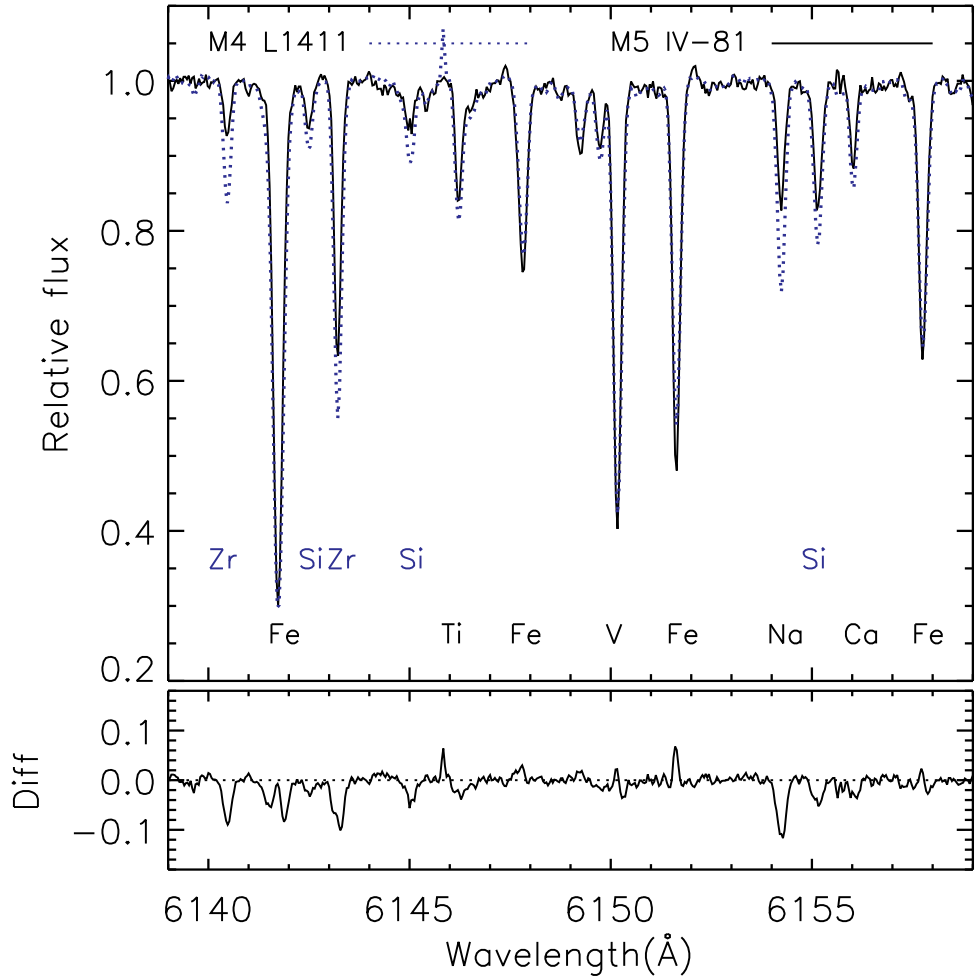}
\caption{The upper panel shows 
observed spectra for M4 L1411 (dotted blue line) and 
M5 IV-81 (solid black line). Various lines within the wavelength 
region are identified. The lower panel shows the difference, 
M4 L1411 $-$ M5 IV-81 after interpolating and rebinning the spectra. 
\label{fig:speccomp1}}
\end{figure}

\begin{figure}
\epsscale{0.8}
\plotone{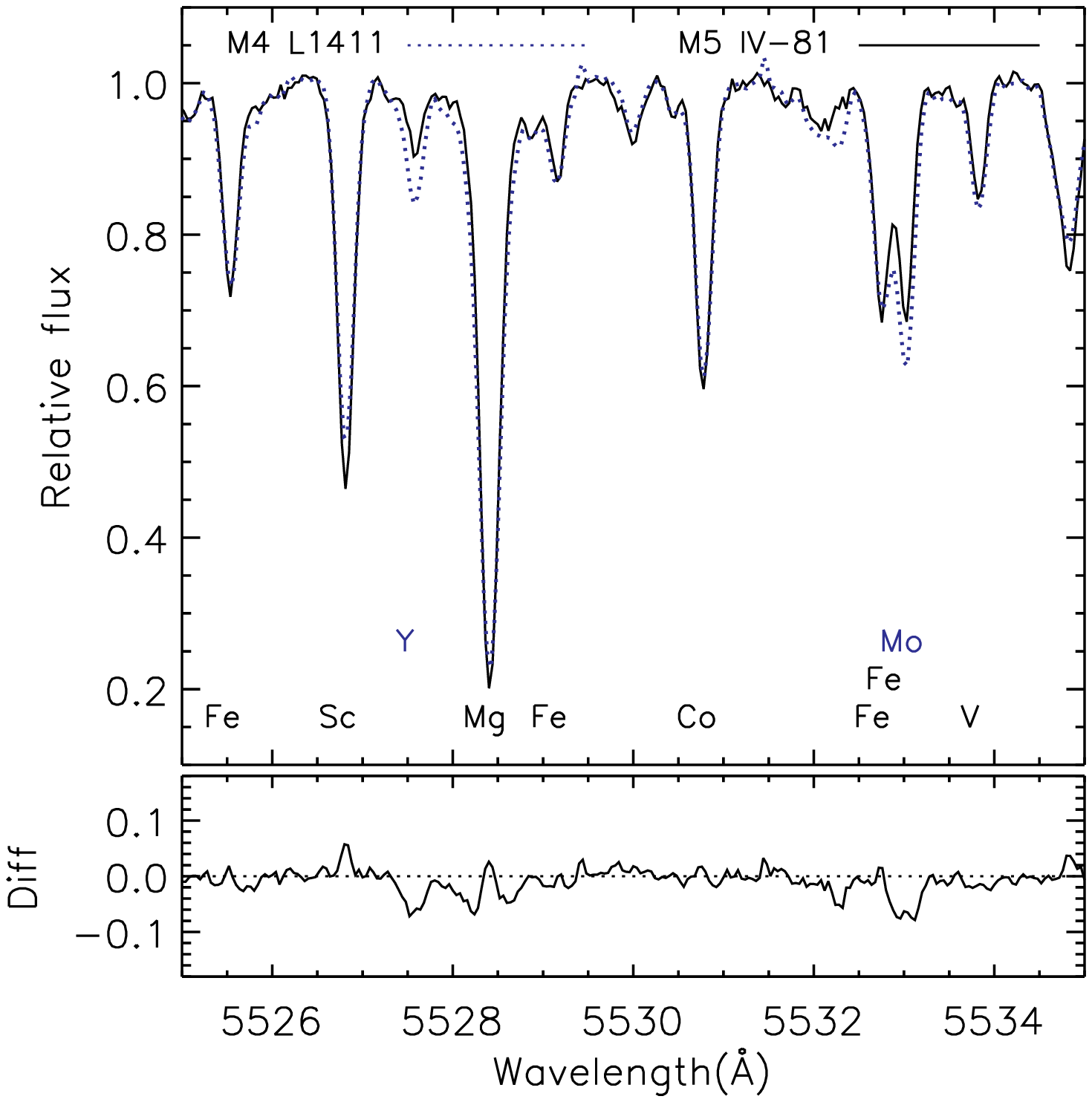}
\caption{Same as Figure \ref{fig:speccomp1} but for a different 
wavelength region. 
\label{fig:speccomp2}}
\end{figure}

\begin{figure}
\epsscale{0.8}
\plotone{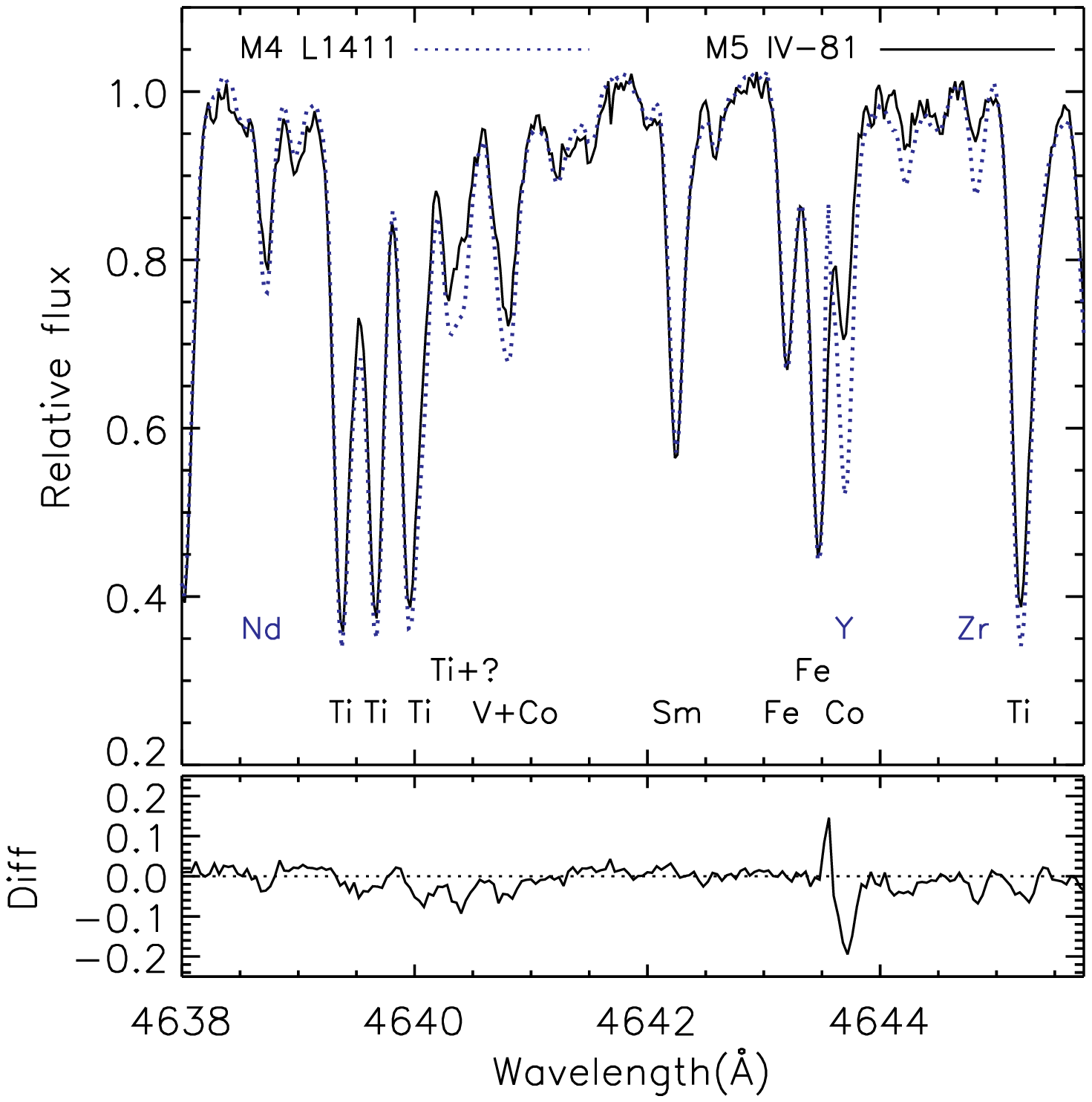}
\caption{Same as Figure \ref{fig:speccomp1} but for a different 
wavelength region.
\label{fig:speccomp3}}
\end{figure}

The sole $r$-process element measured by \citet{M4,M5} is Eu and 
was found to have a slightly higher abundance in M5 relative to M4, 
in contrast to the behavior of the 
$s$-elements. In this study, we have measured the abundances 
for three additional $r$-process elements, Sm, Gd, and Th.  
In Figure \ref{fig:th}, we show synthetic spectra fits to the 
5989\AA\ Th\,{\sc ii} line in which M5 IV-81 has a slightly higher abundance 
than M4 L1411. The 5989\AA\ Th\,{\sc ii} line was also used by 
\citet{aoki07} and Figure \ref{fig:th} shows that in the absence of 
large Th enhancements, abundance estimates are only possible from 
extremely high S/N spectra.  \citet{ivans06} measured 
Th abundances in the $r$-process-rich star HD 221170 
and found that the abundances from 5989\AA\ Th\,{\sc ii} line 
agreed with measurements 
from other Th lines. In Figure \ref{fig:speccomp3}, 
the 4642\AA\ Sm\,{\sc ii} line 
has similar strengths in M4 L1411 and M5 IV-81. 

In Figure \ref{fig:ab2} 
there is a 
trend in which $\Delta$[X/Fe]$_{\rm (M4-M5)}$ 
decreases as the atomic number increases for the $r$-process 
elements Sm, Eu, and Gd. However, this trend does not extend 
to the Th abundances 
(see Figure \ref{fig:ab3}). 
Overall, all four
$r$-process elements, Sm, Eu, Gd, and Th, 
are slightly underabundant in M4 relative to M5, 
$<$$\Delta$[X/Fe]$_{\rm (M4-M5)}$$>$ = $-$0.13 ($\sigma$ = 0.09).

For the neutron-capture elements Pr, Nd, and Hf, the solar 
abundances may be equally attributed 
to the $s$-process 
and $r$-process. Given the behavior of the 
$s$- and $r$-elemental abundances in these clusters, we would therefore expect 
Pr, Nd, and Hf to be overabundant in M4 relative to M5, but that the 
magnitude of the enhancement would lie midway between the bulk of the 
$s$- and $r$- elements. 
Indeed, the abundance differences 
$\Delta$[Pr/Fe]$_{\rm (M4-M5)}$ = +0.09, 
$\Delta$[Nd/Fe]$_{\rm(M4-M5)}$ = +0.12, and 
$\Delta$[Hf/Fe]$_{\rm (M4-M5)}$ = +0.11, 
all lie midway between the heavy $s$-elements, 
$\Delta$$<$[Ba,La,Ce/Fe]$>$$_{\rm(M4-M5)}$ = +0.36, and 
the $r$-process elements
$\Delta$$<$[Sm,Eu,Gd/Fe]$>$$_{\rm(M4-M5)}$ = $-$0.12. 
For Pr, these results 
indicate that either hyperfine structure does not significantly 
affect the abundances within our sample or that fortuitously, 
hyperfine structure 
equally affects the abundances in our sample. 

\section{Discussion}

\subsection{Background}

In terms of chemical composition, globular clusters continue
to present a series of intriguing questions concerning stellar
nucleosynthesis and chemical evolution.
This paper examines the pair of 
clusters - M4 and M5 - and has confirmed and extended the
observations of differences in composition for these clusters of  very
similar iron abundances.
Of the elements examined by us up through the Fe-peak, three are
more abundant in M4 than in M5: Si, Cu, and Zn. Differences in [X/Fe]
are 0.2 to 0.3 dex and might be the same for all three elements.
There are abundance differences for the heavy elements
Rb to Th that range from $+0.6$ (Pb) to $-0.2$ (Gd). 
It is from these abundance differences that
we develop our speculations about stellar nucleosynthesis and
chemical evolution.

Determinations of the compositions of globular cluster stars
suggest the following chemical history for a cluster. The
present population of stars formed from a gas cloud of homogeneous
composition; this postulate is necessary to account for the
fact that within a given cluster all stars have the same
iron abundance (and many other elemental abundances).
Within a cluster, there are star-to-star differences
in the abundances of light elements from C to Al.
Discovery of O, Na, and Al abundance differences among turn-off and subgiant
stars (e.g., \citealt{gratton01})
proves that a cause is to be found in the cluster's
local environment. A possibility, often cited and explored, involves the
selective accretion or pollution of material from the winds of AGB stars,
particularly highly luminous intermediate-mass AGB stars with
hot-bottom convective envelopes capable of nucleosynthesis 
of C to Al  
(e.g., \citealt{cottrell81,denissenkov98,ventura01}), 
although there are quantitative problems with this scenario 
\citep{denissenkov03,fenner04,karakas06}. 
Undoubtedly, an additional contribution to the 
abundance variations involving C, N, Li, and possibly O in red giants
is the presence of mixing into the atmosphere of nuclear-processed material
from the interior. If the convective mixing, expected to a certain degree of
every red giant, is additionally dependent on a stellar property
such as rotation,   star-to-star differences in composition will
result for the red giants.

Given this background, we offer the following idea for the primary
origin of the composition differences between M4 and M5. 
Since the abundances of Si, Cu, Zn, and the heavier elements
do not show a star-to-star spread across the M4 sample, we suppose
that there is no detectable (positive or negative) contribution to their
abundances from the polluting 
stars responsible 
for the
star-to-star abundance variations, and, therefore, the abundance
differences were present between the clouds forming these two
clusters. 

In the case of M5 and many other globular clusters, the
relative abundances (i.e., the [X/Fe] ratios) 
of the elements unaffected by either mixing in
red giants or accretion of pollutants from winds of AGB stars
(or other sources) 
fall within the range exhibited by field stars
of the same iron abundance (e.g., \citealt{gratton04} and 
\citealt{pritzl05}). 
This congruence suggests that most 
globular clusters
form from a cloud that underwent the same pattern of chemical
evolution that led to the majority of the  field halo stars. 

The implications of this suggestion will not be
thoroughly explored here but two points will be mentioned.
One possibility is that field
halo stars were constituents of globular clusters
but some stars, now field stars, may have
been ejected from the present population of clusters. 
Indeed, the globular clusters Pal 5, NGC 5466, and NGC 6712 
exhibit large tidal tails and/or other evidence of severe tidal disruption 
\citep{demarchi99,odenkirchen01,belokurov06}. However, while Pal 5 
shows the globular cluster signature of 
light element abundance variations \citep{smith02}, such 
abundance patterns have yet to be found in field halo stars \citep{gratton00b}. 
Alternatively, the process of cluster formation may have
resulted in the dispersal of stars rather than a stable cluster.
Additionally, the  
suggestion places constraints on the chemical evolution histories of
gas clouds from which clusters and field stars formed. In the case of
metal-poor clusters including M4 and M5 and field stars of comparable
metallicity, chemical evolution was dominated by ejecta from
Type II supernovae; Type Ia supernovae have almost certainly not
made their distinctive contributions to the composition of the
gas cloud. Similarity in [X/Fe] ratios implies similarities in
the mixing of ejecta with ambient gas over the
history of the clouds from which field and cluster stars formed.
For elements 
where [X/Fe] ratios in the ejecta are not very sensitive to
the initial mass function or the initial composition
(i.e., the $\alpha$-elements), different
levels of mixing with pristine gas uncontaminated by X and Fe, 
or mixing with gas previously polluted by ejecta from Type II supernovae 
(same [X/Fe] but different [Fe/H]), 
will give essentially a single value of [X/Fe] at all
[Fe/H]. On the other hand, for elements like Cu where the
yield is dependent on the initial composition of the massive stars,
variations in the mixing processes will lead to a broadening of the
[Cu/Fe] versus [Fe/H] relation.

With this introduction we turn to discussion of the
abundance differences between M4 and M5.

\subsection{The $s$- and $r$-process mix: Ba to Th}

With the usual attribution of Ba to the $s$-process and Eu to the
$r$-process, the higher [Ba/Fe] and the lower [Eu/Fe] for M4 relative to
M5 indicates that the M4 stars formed from gas that
was enriched in $s$-process products  but deficient in $r$-process
products relative to the mixtures in the gas that formed
the M5 stars.
This attribution may be subject to quantitative
analysis with two assumptions.  
It has been  shown that relative abundances of
$r$-process  products in
the interval Ba to Ir in $r$-process enriched stars 
(e.g., \citealt{sneden96,westin00,christlieb04}) 
and in the globular cluster M15 \citep{sneden97,sneden00b,otsuki06} 
closely follows the distribution of $r$-process
products derived from solar system abundances.  
This observation that the $r$-process from Ba to Ir is `universal'
remains without convincing theoretical explanation. Observations
show that universality does not extend to the lighter
elements - Rb to Mo are the observable elements.
Our first assumption  is that the $r$-process contributions to
M4 and M5 are scaled versions of the solar $r$-process
abundances. 
The second assumption is that the $s$-process
contributions are scaled versions of the solar $s$-process
abundances. Resolution of the solar abundances into
$r$- and $s$-process contributions is taken from \citet{simmerer04}. 
These assumptions are not validated by observations
unlike the assumption about the $r$-process. 

\begin{figure}
\epsscale{0.8}
\plotone{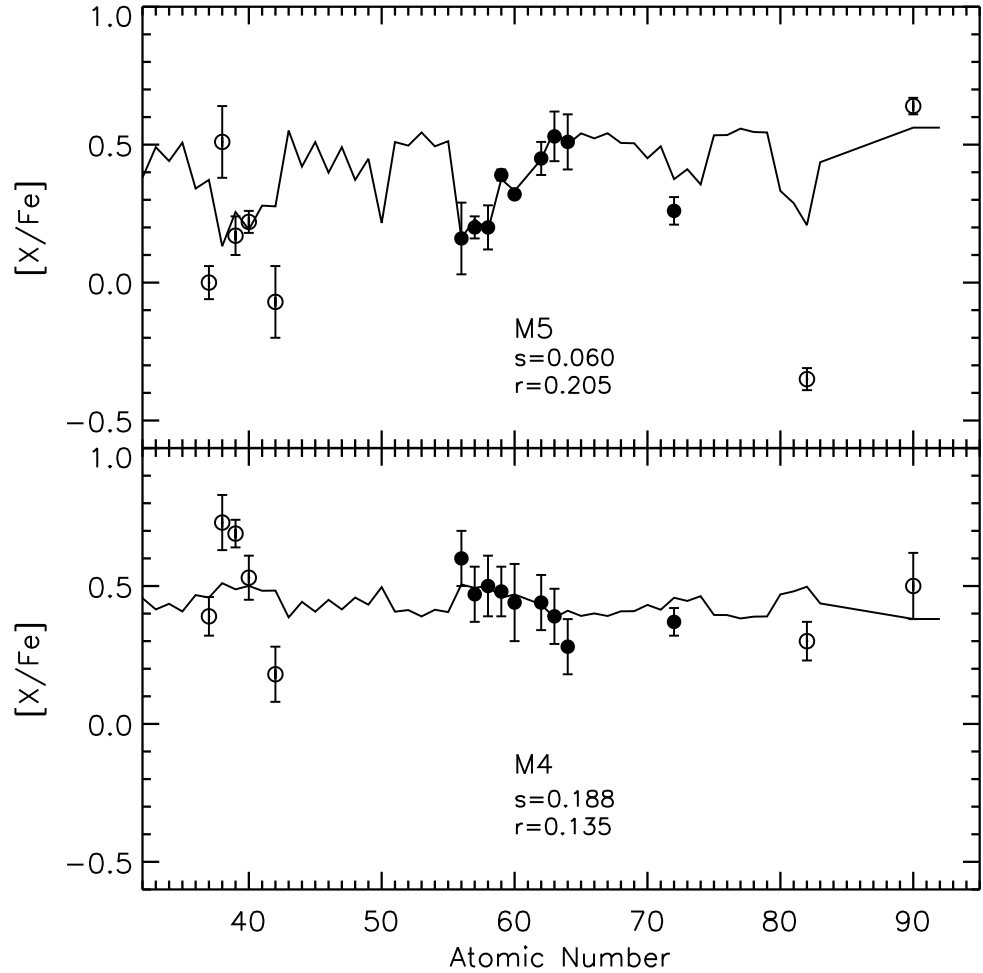}
\caption{Abundance ratios [X/Fe] for Rb to Th in 
M5 (upper) and M4 (lower). In both panels, we show the best fit 
predictions 
to the elements from Ba to Hf (filled circles) 
using scaling factors s and r which were 
multiplied by the 
solar $s$-process and $r$-process abundances respectively.
\label{fig:s2r}}
\end{figure}

For M4 and M5, we adopted their mean abundance ratios [X/Fe]. 
We then took linear combinations of the solar $s$- and $r$-process 
abundances (adopting [Fe/H] = $-$1.25) exploring the full range of 
parameter space (i.e., considering values from 0.01 to 1.00 times 
the solar $s$- and $r$-process abundances). 
For each combination of $s$- and $r$-process 
abundances, we compared the predicted abundance ratios with the 
observed abundance ratios. For each cluster, we located the optimal 
scaling factors for the $s$- and $r$-process. When considering the elements 
from Ba to Hf, we find s = 0.188 and r = 0.135 
for M4 and s = 0.060 and r = 0.205 
for M5 where `s' and `r' are the scaling factors 
of the solar $s$- and $r$-process abundances respectively. 
For both clusters, the predicted and 
measured abundances are in very good agreement 
(see Figure \ref{fig:s2r}). 
For M4 we find $\Delta$[X/Fe]$_{\rm predicted - observed}$ 
= 0.04 ($\sigma$ = 0.05)
and for M5 we find $\Delta$[X/Fe] = 0.02 ($\sigma$ = 0.04). 

For both clusters, the abundances of Th lie above the predictions, albeit 
by similar amounts. 
The ratio of Th to Eu is identical in both clusters, [Th/Eu] = 0.11. 
Since Th and Eu are produced exclusively by the $r$-process, 
the identical [Th/Eu] ratios in M4 and M5 indicate that the 
$r$-process universality extends to Th in both clusters. Furthermore, 
the identical [Th/Eu] ratios suggests that no differential decay of Th has 
occurred and that these two clusters have essentially identical ages. 
(An uncertainty of 0.1 dex in the [Th/Eu] ratio would result in an 
age uncertainty of roughly 3 billion years.) 

For both clusters, the observations of Pb fall below the predictions 
with M4 lying closer to the prediction than M5 owing to the 
different abundances, [Pb/Fe] = 0.30 in M4 and 
[Pb/Fe] = $-$0.35 in M5. As mentioned, the Pb abundance in M5 is typical 
of field halo stars at the clusters' metallicities. 
The cluster-to-cluster differences between the observed and predicted Pb 
abundances indicate that the $s$-process similarities break down 
for Pb. In the Sun, more than 50\% of $^{208}$Pb, the most abundant 
isotope, is due to the so-called "strong component" 
(e.g., \citealt{clayton67}, \citealt{kappeler89}), where low mass 
metal-poor AGB stars are the likely site of this component. 
Therefore, the differences between the observed and predicted Pb abundances 
in M4 and M5 
provide a hint that the sources of the $s$-process, in particular 
the strong component, differ between these clusters. 

Had we included all elements from Ba to Th 
in determining the scaling factors for the solar $s$- and $r$-process 
abundances, we would have obtained very similar values 
(M4: s = 0.175, 
r = 0.138, and M5: s = 0.053, r = 0.210) 
as obtained when fitting the Ba to Hf abundances. 
These new predictions 
again provide a poor fit to Pb and Th. 

The mix of $r$- and $s$-process contributions for M5 is typical of
the mix found in halo field stars but the mix for M4 is quite
atypical. 
This is well shown when considering the ratio of the La to Eu 
abundances. Inspection of Figure 7 in \citet{simmerer04} shows that 
M5, $\log\epsilon$(La/Eu) = 0.28, is representative of field stars 
at the cluster's metallicity. 
However, M4, $\log\epsilon$(La/Eu) = 0.70, 
is in the far tail of the distribution of La/Eu ratios for
field stars and close to the stars labelled as $s$-process enriched. 
Yet, as our fits to the abundance patterns show, the $s$-process
in M4 has the same pattern of abundances as in M5 and the Sun. 
Thus, no new site for $s$-process operation is needed to account for the
M4-M5 differences, at least for the elements Ba to Hf. 
We presume that AGB stars are the active site. 
Yet, the difference in $s$-process pollution  between M4 and M5
might suggest that M4 originated in a different part of the
Galaxy to M5 and local halo stars.

\subsection{Cu to Zr}

The successful dissection of the Ba-Ir abundances into different
mixing fractions of solar-like $r$- and $s$-process abundance
patterns cannot be easily extended to lighter
elements in the Cu-Zr interval. 
This is likely attributable to the appearance of an additional
one or more processes of stellar nucleosynthesis.

One expects the weak $s$-process to contribute to the elements 
Cu to Zr as revealed from the fit to the solar
abundances (e.g., \citealt{kappeler89}). 
This neutron capture process occurs in
massive stars in the He-burning and the C-burning shells with
products ejected without substantial modification in the
supernova explosion. At initial metallicities [Fe/H] $\sim -1$,
the contributions from the He-burning shell appear to be
more important than from the C-burning shell. This $s$-process
does not provide significant yields beyond the cross-section
bottleneck at the neutron magic number $N=50$ and, thus, the
Ba-Ir interval is unaffected by this contribution from massive
stars.  Predicted yields
for this weak $s$-process are sketched by \citet{pignatari07}. 
It is also below [Fe/H] $\sim -1$ that the weak
$s$-process may become so ineffective that explosive
Si-burning may replace it as the source of Cu, Zn, and other
weak $s$-process products. 

The $r$-process, whose site is probably the deep interior of
Type II supernovae from massive stars, is likely to contribute also 
to the Cu-Zr interval, in part or in whole. Unfortunately, there
are not even qualitative predictions about these $r$-process
yields. Abundance analyses of very $r$-process enriched stars
by Sneden and colleagues have shown a star-to-star variation in
these lighter elements --  Ga to Ag --  in their abundance ratios
with respect to the abundances of the heavier elements from Ba and up.
These variations have been ascribed to `a second $r$-process' but
may at least in part reflect a weak $s$-process contribution from massive
stars, also the likely site of the $r$-process(es). 
(Recall that we
have discounted the possibility that Type Ia supernovae are
contributing to the chemical evolution of M4 and M5 and field
stars of comparable or lower metallicity.)

As expected, Figure \ref{fig:s2r} shows that predictions based on 
the scaling factors for the solar $s$- and $r$-process abundances 
that match the elements from Ba to Hf provide a poor fit to the 
elements from Rb to Mo. Had we included all elements from Rb to Th 
in the fit, the scaling factors would be essentially unchanged 
(M4: s = 0.060, r = 0.198, and M5: s = 0.188, r = 0.138), and 
the fit to the elements Rb to Mo would remain poor. 

\subsection{The case of Si}

An overabundance of $\alpha$-elements is commonly taken as
the signature of gas contaminated to a major degree by ejecta of
Type II supernovae. Here, the $\alpha$-element directory includes O,
Mg, Si, S, and Ca with Ti as an honorary member according
to observers. Ca and Ti have very similar abundances
in M4 and M5, but Si is definitely overabundant in M4 with
respect to M5 (Figure \ref{fig:ab1} and Table \ref{tab:abund1}). 
In M4, the overabundance of Mg, 
[Mg/Fe] = 0.56 ($\sigma$ = 0.05) from 12 stars \citep{rbpbm4m5}, 
is similar to that of Si, [Si/Fe] = 0.58 ($\sigma$ = 0.08). 
This contradicts findings by \citet{M5} who 
found that stars in M4 have Si abundances about 0.2 dex higher than 
Mg. A more careful investigation of the difference in [Mg/Si] ratios 
in M4 by different authors would need to account for systematic 
differences. 

Both clusters exhibit the classic signature of $\alpha$-elements in
metal-poor stars -- [$\alpha$/Fe] $\simeq$ 0.3 dex --
but this signature is different in its details, as just noted. 
There is no obvious explanation for the enhanced Si abundances in M4. 
Si and Ca (as well as S and Ar) are all produced by incomplete 
explosive Si burning and explosive O burning in massive stars. Therefore, 
any Si excess in M4 relative to M5 should be accompanied by a similar 
Ca excess. 

\subsection{Comments on Cu and Zn nucleosynthesis}

Pioneering measurements of Cu and Zn in field stars showed that 
between $-$2.5 $\le$ [Fe/H] $\le$ 0.0, [Cu/Fe] declined with decreasing 
metallicity whereas [Zn/Fe] was constant and solar 
\citep{peterson81,gratton88,sneden88,sneden91}. 
Although a large amount of data are now available for these elements,
the nucleosynthetic origins of Cu and Zn remain unclear with 
both elements likely requiring multiple production sites and 
metallicity dependent yields. For Cu, Type Ia supernovae 
\citep{matteucci93,mishenina02} and massive stars 
\citep{timmes95,romano07} are proposed to be the dominant producers with 
low mass AGB stars believed to provide only 
a minor contribution \citep{matteucci93}. 
For Zn, massive stars \citep{timmes95} and Type Ia supernovae 
\citep{matteucci93} have been suggested as being the major 
producers. Other potential sources of Zn include 
the weak $s$-process in massive stars and the 
main $s$-process in low mass AGB stars \citep{matteucci93}, 
and at very low metallicity, the $s$-process or $r$-process 
in massive stars \citep{umeda02,heger02}. 

The abundances measured in M4 and M5 offer additional insight into the 
production sites of Cu and Zn. These metal-poor globular clusters 
were formed early in the history of our Galaxy, before Type Ia 
supernovae contributed to their chemical evolution. Therefore, 
our measurements confirm that stars other than 
Type Ia supernovae must produce significant amounts of Cu and Zn 
at low metallicities. 
Abundance differences of roughly 0.3 dex 
are found for the $\alpha$-element Si (produced primarily in massive stars) 
as well as the $s$-process elements (produced in AGB stars or 
massive stars). In light of the abundance differences for Cu and Zn 
in these two clusters, we now examine the synthesis of 
Cu and Zn in massive stars and AGB stars based on recent theoretical 
models. 

\subsubsection{Massive star production}

Both the stable isotopic nuclei of Cu and the five of Zn are produced, 
at low metallicities ([Fe/H] $<$ $-$1), only by the complete explosive 
Si burning, i.e., in the deepest regions of the exploding star where 
the peak temperature exceeds 5 $\times$ 10$^9$ K 
and therefore matter can reach the 
nuclear statistical equilibrium 
\citep{chieffi04}. 
The striking similarities of the relative abundances of the elements 
between Ca and Ni seem to clearly indicate that the core collapse 
supernovae responsible for the bulk production of the nuclei in this 
range are very similar in M4 and M5. In this respect Cu and Zn 
should also behave similarly (because of their production site) but there 
is a "caveat": they are less abundant (relative to hydrogen) 
than Co and Ni and hence they 
would be the first elements within the Fe peak nuclei to be modified by 
the passage of a neutron flux. 

In massive stars, He- and C-burning can lead to activation of the 
$^{22}$Ne($\alpha$,$n$)$^{25}$Mg neutron source whose high neutron 
densities are responsible for synthesizing $s$-process elements 
up to Sr \citep{busso99}. 
If the weak $s$-process is solely responsible for the enhanced Cu and Zn 
in M4, we would expect the abundances of 
Rb relative to nearby elements such as Sr, Y, 
and Zr to also differ between these clusters 
\citep{tomkin83,lambert95,abia01}. Observation support for this scenario 
comes from enhanced Rb abundances in luminous Galactic OH/IR stars of 
solar metallicity 
\citep{gh06}
that are matched by theoretical models 
\citep{vanraai08}. However, \citet{rbpbm4m5} showed 
that in M4 and M5 the ratio [Rb/Zr] was identical within the measurement 
uncertainties and therefore, it may be difficult to ascribe any 
difference in Cu and Zn solely to the weak $s$-process, unless the 
production of [Rb/Zr] has a metallicity dependence and is not significant 
at low metallicity. As discussed above, no comprehensive 
predictions are presently available. 

\subsubsection{AGB production}

Models presented in \citet{karakas08} indicate that Zn, in general, 
is not altered by AGB nucleosynthesis. Zn can be produced in the most 
massive, lowest metallicity AGB stars, with the largest abundance 
predicted to be [Zn/Fe] $\sim 0.4$~dex in a 5$M_{\odot}$, $Z=0.004$ model. 
The total expelled mass of Zn in this case is small ($\sim 10^{-6}M_{\odot}$).
Preliminary results for Cu suggest that neutron-capture nucleosynthesis in 
intermediate-mass low-metallicity AGB stars could be an extra production 
site for this element. Both stable isotopes of Cu are predicted to be 
enhanced in the He-shell of 4 to 8$M_{\odot}$ AGB stars, with the 
final elemental Cu surface abundance estimated to be [Cu/Fe] $\sim 0.8$~dex 
in a 5$M_{\odot}$, $Z = 0.004$ model. 

The results for Zn agree with the \citet{travaglio04} 
predictions that only 2.4\% of 
elemental Zn comes from the $s$ process, however the results for Cu are 
less clear, where \citet{travaglio04} 
estimate that 5.2\% of Cu is produced in AGB stars. We would need to include
updated 
AGB yields of Cu and Zn in a chemical evolution model to study the 
consistency with the \citet{travaglio04} results, as well as to 
determine just how efficient intermediate-mass AGB production really is, 
along with an exhaustive study of the many uncertainties affecting the
AGB nucleosynthesis models. For example, 
one of the largest uncertainties is the mass-loss law used in the 
calculations, as this determines the number of third dredge-up mixing 
episodes. If the number of mixing episodes were to be halved, then about 
four times less Cu would be produced in the 5 M$_{\odot}$ example given 
above. 

\begin{figure}
\epsscale{0.8}
\plotone{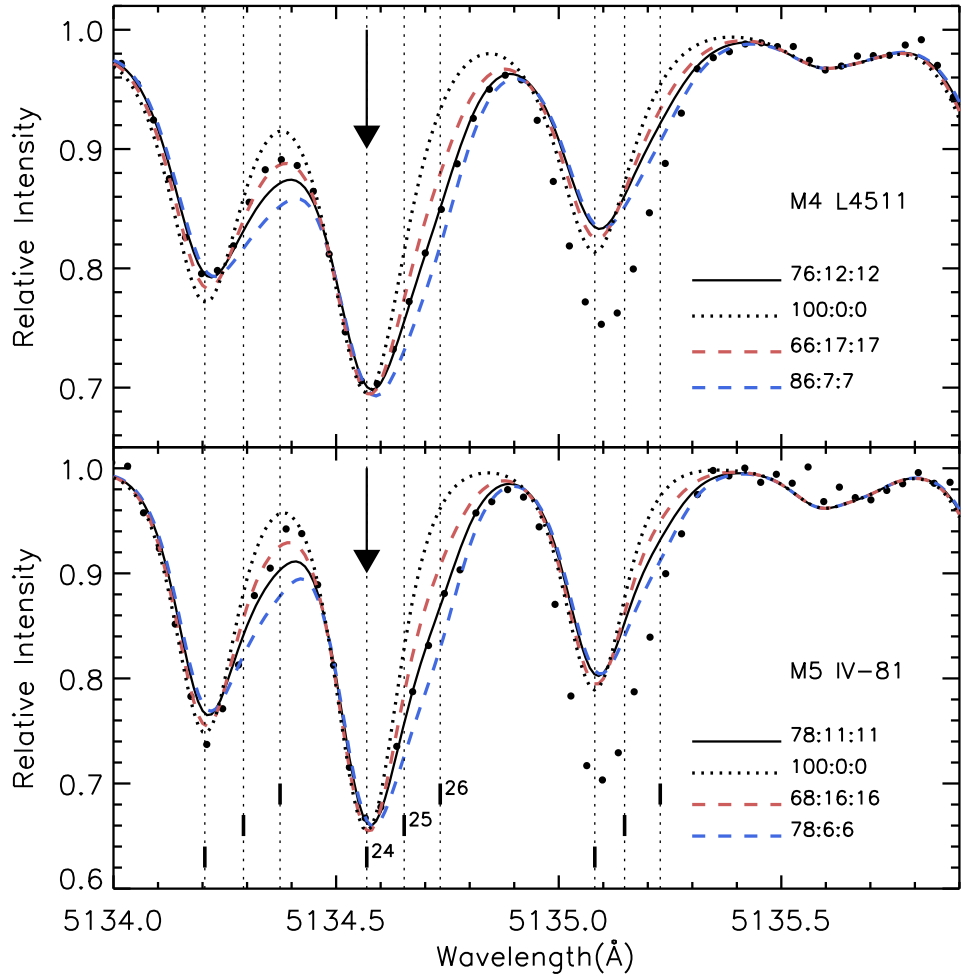}
\caption{Spectra of M4 L4511 (upper) and M5 IV-81 
near the 5134.6\AA\ MgH line. 
The positions of $^{24}$MgH, $^{25}$MgH, and $^{26}$MgH are marked 
by dashed lines. The filled circles represent the observed spectra, 
the best fit is shown by the solid line, and unsatisfactory ratios 
are also shown.
\label{fig:mgiso}}
\end{figure}

If 5$M_\odot$ metal-poor AGB stars are solely 
responsible for the enhanced Cu and Zn 
in M4, we would expect differences in the abundances of the heavy Mg 
isotopes between M4 and M5 
\citep{karakas03,karakas06b}. Observations of high $^{26}$Mg/$^{24}$Mg 
ratios in field and cluster stars confirm the AGB yields 
\citep{gl2000,6752,mghdwarf}. Although the resolution is less than ideal 
(R = 55,000 in this study compared with R $\ge$ 90,000 in our previous 
studies of Mg isotope ratios in giant stars in globular clusters), 
we have measured the Mg isotope ratios using the 
same methods as \citet{6752,mghsubaru}. The ratios are given in
Table \ref{tab:mgiso} and examples of synthetic spectra fits are 
shown in Figure \ref{fig:mgiso}. Given the 
resolution, we adopt $^{25}$Mg = $^{26}$Mg in the reported ratios, 
and therefore we regard these results as preliminary because our analyses 
of stars in other globular 
clusters have showed that $^{25}$Mg and $^{26}$Mg can 
behave independently. The uncertainties in the ratios are 
$b$ $\pm$ 5 and $c$ $\pm$ 5 when expressing the ratio as 
$^{24}$Mg:$^{25}$Mg:$^{26}$Mg = ($100-b-c$):$b$:$c$. 
Nevertheless, these preliminary 
results provide important additional information into the chemical 
evolution of these clusters. Within M4 and M5, the Mg isotope 
ratios are constant, in contrast to what is seen in the more 
metal-poor globular clusters 
NGC 6752 and M13 \citep{shetrone96b,6752,mghsubaru}. 
Given the small amplitude 
[Al/Fe] variation in our sample ($\Delta$[Al/Fe]$_{\rm M4}$ = 0.23 
compared with $\Delta$[Al/Fe]$_{\rm NGC~6752}$ = 1.06), it is not 
surprising that the Mg isotope ratios show a smaller variation in M4 than 
in NGC 6752 (and M13). 
In both M4 and M5, the ratios are very similar with a 
typical cluster ratio comparable to the solar value, 
$^{24}$Mg:$^{25}$Mg:$^{26}$Mg = 80:10:10. Even at our resolution, 
the isotope ratios $^{25}$Mg/$^{24}$Mg and $^{26}$Mg/$^{24}$Mg
in M4 and M5 exceed values measured in field 
halo stars at the clusters' metallicities \citep{gl2000,mghdwarf,melendez07}
and predictions from chemical evolution models \citep{alc01,fenner03}. 
For example, the halo 
dwarf Gmb 1830 has [Fe/H] = $-$1.30 and 
$^{24}$Mg:$^{25}$Mg:$^{26}$Mg = 94:3:3 \citep{tl80}. 
Therefore, the clouds from which M4 and M5 formed were not like the 
clouds from which halo stars like Gmb 1830 formed. 
In summary, the Mg isotope ratios in M4 and M5 do not support the 
scenario in which intermediate AGB stars are responsible for the 
Cu and Zn abundance differences between these clusters. For both 
clusters the isotope ratios exceed the values found in field stars at 
the same metallicity. 

Further constraints on the AGB contribution to the Cu 
abundance differences in M4 and M5 come from the most massive 
globular cluster, $\omega$ Centauri. In $\omega$ Cen, 
\citet{cunha02} found a roughly constant ratio  
[Cu/Fe] $\simeq$ $-$0.5 as [Fe/H] ranged from $-$2.0 to $-$0.8 
in their sample of 40 stars. The same sample was previously analyzed 
by \citet{norris95b} who found that the abundance ratios [X/Fe] 
for $s$-process elements increased as [Fe/H] increases. 
Therefore, the low-mass AGB stars presumably responsible for the 
increasing [$s$-element/Fe] as [Fe/H] increases do not synthesize
Cu. However, 
$\omega$ Cen exhibits 
a number of traits not found in other globular clusters. 
Detailed 
chemical evolution models such as those performed by 
\citet{fenner04} and \citet{renda04} will help unravel 
the origin of the Cu and Zn abundance differences in M4 and M5 
as well as the 
synthesis of Cu and Zn in metal-poor stars. 

\subsection{Physical and kinematic differences between M4 and M5}

Finally, we briefly consider the physical and kinematic parameters 
for M4 and M5. 
Within the Galactic globular clusters, M4 has a typical luminosity 
(i.e., mass), $M_V$ = $-$7.20, 
while M5 is one of the more luminous (i.e., massive) clusters, 
$M_V$ = $-$8.81 \citep[Feb.\ 2003 web version]{harris96}. 
The space velocities for these two clusters are considerably 
different. \citet{dinescu99} have shown that 
M5 has a very large apocentric radius ($R_a$ = 35.4 kpc) 
while M4 has a very small apocentric radius ($R_a$ = 5.9 kpc). 
Given that M4's orbit is confined to the 
inner disk and bulge, it would be of great interest to measure 
the abundances of Si, Cu, Zn, and $s$-process elements in a sample of 
inner disk and bulge stars at the metallicity of M4. Perhaps 
the elevated abundances of Si, Cu, Zn, and $s$-elements in M4 are 
representative of stars born at small Galactocentric radii. 
However, 
we note that \citet{fulbright07} measured the abundances of Si, Ca, and 
Ti in bulge giants including four with $-$1.5 $\le$ [Fe/H] $\le$ $-$1.0. All 
have [Si/Fe] ratios comparable to [Ca/Fe] and [Ti/Fe]. 
Two stars below [Fe/H] = $-$1 in the \citet{fulbright07} sample 
have high [Al/Fe] and [Na/Fe] along with 
low [O/Fe], the abundance signature of hydrogen burning at high temperatures 
seen in globular clusters. 
Indeed, they speculate that these two stars may be members of 
the bulge globular cluster NGC 6522. 
Abundances of $s$-process elements were measured by 
\citet{mcwilliam94} in three bulge giants 
with [Fe/H] $\le$ $-$0.8, none of which showed $s$-process enrichments. 
Therefore, the enhanced abundances of Si and $s$-elements in M4 may 
not be representative of stars born at small Galactocentric radii. 

\section{Concluding remarks}
\label{sec:summary}

In this paper we present abundance ratios [X/Fe] for a large number 
of $\alpha$-, Fe-peak, $s$-process, and $r$-process elements for 12
bright giants in the globular cluster M4 and 2 bright giants 
of the globular cluster M5. This comprehensive 
abundance analysis is only possible due to the large wavelength coverage, 
high resolution,  
and very high S/N spectra. For all elements in this study, 
we find no evidence for star-to-star abundance variations in either 
cluster. We confirm and extend upon previous results for these 
clusters by showing that 
(1) for the elements from Ca to Ni, M4 and M5 have identical 
abundance ratios, (2) M4 shows 
overabundances by roughly 0.3 dex for Si, Cu, Zn, and all 
$s$-process elements relative to M5, and (3) for the $r$-process elements, 
M5 may have slightly higher abundances than M4 by 0.1 dex. 

We also measure 
Mg isotope ratios and find that the ratios are solar in both clusters, 
with no sign of any star-to-star variation within each cluster. 
The ratios $^{25}$Mg/$^{24}$Mg and $^{26}$Mg/$^{24}$Mg exceed values 
found in field halo stars at the same metallicity, e.g., Gmb 1830, 
which implies differences in the clouds from which globular clusters 
and field halo stars formed. 
However, we regard these ratios as preliminary since 
the spectral resolution was insufficient to accurately distinguish 
$^{25}$Mg from $^{26}$Mg. 

There is no clear explanation for the M4-M5 Si abundance differences 
since the abundances of Ca should, but do not, follow the behavior of Si. 
For the elements from Ba to Hf, we find that the mean abundances in M4 and M5 
are well explained by scaled versions of the solar $s$- and $r$-process 
abundances, albeit with different mixes of $s$- and $r$-process material 
for each cluster. Therefore, no new $s$-process site is required to 
explain the M4-M5 abundance differences for the elements from Ba to Hf. 
However, although the Th abundances lie above these predictions, 
the ratio [Th/Eu] is identical in both clusters indicating that the 
universality of the $r$-process extends to Th in these clusters and that 
no differential decay of Th has occurred, i.e., the clusters have 
identical ages. The Pb abundances lie below the predictions, by 
different amounts for each cluster. Therefore, the sources of 
the $s$-process may differ between M4 and M5, at least regarding 
the production of Pb via the strong component. 

The abundance differences between M4 and M5 
for Cu and Zn are particularly intriguing given 
that their nucleosynthetic origins continue to be debated. 
The $s$-process elements, produced in 
AGB stars (and massive stars), share a similar abundance behavior 
to Cu and Zn in M4 and M5. Updated, but preliminary, yields from 
AGB models indicate that small amounts of Zn may be produced only 
in the most massive AGB stars. These models also predict that 
the most massive AGB stars may produce Cu in contrast to our 
current understanding of Cu production. Massive AGB 
stars are expected to produce large amounts of the neutron-rich 
Mg isotopes, and observations of high $^{26}$Mg/$^{24}$Mg 
ratios in field and cluster stars confirm the AGB yields. However, 
preliminary measurements 
show no difference in the Mg isotope ratios between these two clusters 
which constrains the contribution of intermediate-mass 
AGB stars to the Cu and Zn 
enhancements in M4. 
Si, which is produced in massive stars, shows a similar abundance 
behavior to Cu and Zn 
in M4 and M5. Cu and Zn may be produced in massive stars 
via the weak $s$-process. While the abundance ratios [Rb/Sr], 
[Rb/Y], and [Rb/Zr] are predicted to increase via the weak $s$-process, our 
measurements do not reveal any cluster-to-cluster variations in 
these abundance ratios which suggests that either 
massive stars are not responsible for the Cu and Zn differences 
or that metal-poor massive stars do not alter the [Rb/Zr] ratio. 
Of great interest would be detailed chemical evolution modeling 
of these two clusters to 
gain insight into the origin of the Cu and Zn abundance differences 
and therefore their nucleosynthesis production sites. 

\acknowledgments

This research has made use of the SIMBAD database,
operated at CDS, Strasbourg, France and
NASA's Astrophysics Data System. DY thanks 
Inese Ivans and John Norris for valuable discussions and the 
referee for helpful comments. 
AIK acknowledges support from the Australian Research Council's 
Discovery Projects funding scheme (project number DP0664105), and 
thanks Maria Lugaro for help in developing the extended 
nucleosynthesis network. 
DLL thanks the Robert A.\ Welch Foundation of Houston for support. 
This research was 
supported in part by NASA through the American Astronomical Society's Small 
Research Grant Program.

\clearpage

\begin{deluxetable}{lcccc} 
\tabletypesize{\footnotesize}
\tablecolumns{5} 
\tablewidth{0pc} 
\tablecaption{Stellar parameters.\label{tab:param}}
\tablehead{ 
\colhead{Star} &
\colhead{\teff} &
\colhead{log $g$} &
\colhead{$\xi_t$} &
\colhead{[Fe/H]} 
}
\startdata
M4 L1411  & 4025 & 0.80 & 1.75 & $-$1.23 \\
M4 L1501  & 4175 & 1.00 & 1.55 & $-$1.29 \\
M4 L1514  & 3950 & 0.30 & 1.85 & $-$1.22 \\
M4 L2307  & 4125 & 0.95 & 1.65 & $-$1.19 \\
M4 L2406  & 4150 & 0.15 & 2.20 & $-$1.30 \\
M4 L2617  & 4275 & 1.25 & 1.65 & $-$1.20 \\
M4 L3209  & 4075 & 0.75 & 1.95 & $-$1.25 \\
M4 L3413  & 4225 & 1.10 & 1.75 & $-$1.23 \\
M4 L3624  & 4225 & 1.05 & 1.60 & $-$1.29 \\
M4 L4511  & 4150 & 1.05 & 1.70 & $-$1.22 \\
M4 L4611  & 3925 & 0.15 & 1.45 & $-$1.09 \\
M4 L4613  & 3900 & 0.20 & 1.70 & $-$1.25 \\
M5 IV-81  & 4050 & 0.30 & 1.90 & $-$1.28 \\
M5 IV-82  & 4400 & 1.20 & 1.75 & $-$1.33 \\
\enddata

\end{deluxetable}

\clearpage
\pagestyle{empty}
\setlength{\voffset}{24mm}
\begin{deluxetable}{lccrccccccccccccccc} 
\tabletypesize{\tiny}
\rotate
\tablecolumns{19} 
\tablewidth{0pc} 
\tablecaption{Equivalent widths [ONLINE ONLY].\label{tab:ew}}
\tablehead{ 
\colhead{Wavelength (\AA)} &
\colhead{Species} &
\colhead{$\chi$ (eV)} &
\colhead{$\log gf$} &
\colhead{Source $gf$} &
\colhead{M4 L1411} &
\colhead{M4 L1501} &
\colhead{M4 L1514} &
\colhead{M4 L2307} &
\colhead{M4 L2406} &
\colhead{M4 L2617} &
\colhead{M4 L3209} &
\colhead{M4 L3413} &
\colhead{M4 L3624} &
\colhead{M4 L4511} &
\colhead{M4 L4611} &
\colhead{M4 L4613} &
\colhead{M5 IV-81} &
\colhead{M5 IV-82} 
}
\startdata
5665.554 & 14.0 & 4.92 & $-$2.04 & RC02 & 49.2 & 46.6 & 44.3 & 50.1 & 55.7 & 40.6 & 48.8 & 43.1 & 46.5 & 50.4 & 51.3 & 45.2 & 35.1 & \ldots \\
5690.430 & 14.0 & 4.93 & $-$1.83 & RC02 & 48.4 & 49.1 & 48.8 & 49.4 & 58.4 & 51.1 & 47.5 & 47.3 & 49.0 & 50.3 & 47.1 & 46.0 & 36.4 & 30.9 \\
5948.550 & 14.0 & 5.08 & $-$1.23 & RC02 & 76.3 & 80.3 & 77.0 & 82.1 & 100.5 & 87.6 & 75.7 & 77.9 & \ldots & 78.5 & \ldots & \ldots & 71.6 & 61.9 \\
6142.490 & 14.0 & 5.62 & $-$1.48 & IK01 & 21.2 & 22.5 & \ldots & 23.5 & \ldots & 22.5 & 21.3 & 23.2 & 23.8 & 22.8 & 18.8 & 20.6 & 14.0 & 12.9 \\
6145.020 & 14.0 & 5.61 & $-$1.37 & RC02 & 29.7 & 31.9 & \ldots & 31.8 & 35.6 & 31.2 & 30.7 & 28.4 & 29.7 & 32.3 & 29.4 & 30.8 & \ldots & 20.6 \\
6155.134 & 14.0 & 5.62 & $-$0.76 & RC02 & 57.0 & 60.3 & 54.7 & 60.2 & 69.2 & 63.0 & 57.7 & 57.2 & 61.5 & 61.0 & 56.9 & 55.3 & 42.1 & 42.4 \\
6721.840 & 14.0 & 5.86 & $-$0.94 & RC02 & 29.6 & 35.8 & 34.2 & 34.2 & 38.3 & 35.4 & 31.5 & 27.3 & 31.7 & 35.8 & 43.6 & 42.1 & 20.6 & 17.8 \\
5581.980 & 20.0 & 2.52 & $-$0.56 & LUCK & \ldots & 122.8 & \ldots & \ldots & \ldots & 123.3 & \ldots & \ldots & 121.0 & \ldots & \ldots & \ldots & \ldots & 104.7 \\
6166.440 & 20.0 & 2.52 & $-$1.14 & RT03 & 122.6 & 98.9 & \ldots & 110.8 & 95.6 & 101.6 & 120.8 & 103.7 & 98.8 & 109.8 & \ldots & \ldots & 111.5 & 77.0 \\
6455.600 & 20.0 & 2.52 & $-$1.29 & IK01 & 107.1 & 88.1 & 112.5 & 100.3 & 79.1 & 88.0 & 104.8 & 87.4 & 84.5 & 93.1 & 117.9 & 114.0 & 98.8 & 64.2 \\
6499.650 & 20.0 & 2.52 & $-$0.82 & IK01 & \ldots & 120.6 & \ldots & \ldots & 122.4 & 119.8 & \ldots & 123.0 & 117.8 & \ldots & \ldots & \ldots & \ldots & 100.2 \\
5526.820 & 21.1 & 1.77 & 0.13 & PN00 & 113.9 & 103.7 & 114.9 & 109.8 & \ldots & 98.1 & 111.6 & 109.3 & 102.7 & 109.0 & 118.1 & 112.5 & 119.8 & 106.5 \\
6604.599 & 21.1 & 1.36 & $-$1.48 & PN00 & 81.3 & 72.2 & 88.0 & 78.4 & 82.9 & 71.6 & 81.9 & 71.1 & 69.2 & 77.4 & 90.8 & 87.4 & 87.1 & 64.3 \\
5648.567 & 22.0 & 2.49 & $-$0.25 & RC02 & 61.6 & 38.2 & 66.8 & 49.6 & 29.2 & 36.4 & 61.1 & 41.8 & 36.3 & 44.2 & 73.5 & 71.9 & 51.6 & 19.0 \\
5689.490 & 22.0 & 2.30 & $-$0.47 & RC02 & \ldots & 43.0 & \ldots & \ldots & \ldots & 45.9 & \ldots & 48.2 & 42.1 & 53.3 & \ldots & \ldots & 57.1 & 18.6 \\
5716.460 & 22.0 & 2.30 & $-$0.70 & RC02 & 46.7 & 27.7 & 53.8 & 38.1 & 19.8 & 26.6 & 47.6 & 29.5 & 26.7 & 32.6 & 64.8 & 60.6 & 38.0 & 11.3 \\
5720.480 & 22.0 & 2.29 & $-$0.90 & RC02 & 38.2 & 21.9 & 45.0 & 30.6 & 15.7 & 22.4 & 38.3 & 24.8 & 26.3 & 25.5 & 55.2 & 52.1 & 31.5 & 9.0 \\
5739.464 & 22.0 & 2.25 & $-$0.60 & RC02 & 54.4 & 32.4 & 64.1 & 45.1 & 25.2 & 31.9 & 56.8 & 37.1 & 31.9 & 38.5 & \ldots & \ldots & 47.1 & 14.7 \\
5739.982 & 22.0 & 2.24 & $-$0.67 & RC02 & 47.1 & 28.5 & 54.0 & 40.1 & 24.9 & 26.2 & 41.7 & 30.2 & 26.0 & 34.0 & 64.5 & 62.5 & 35.0 & 11.3 \\
5903.317 & 22.0 & 1.07 & $-$2.14 & RC02 & 91.5 & 52.3 & 106.9 & 72.3 & 46.7 & 50.0 & 89.0 & 57.2 & 49.4 & 61.0 & 112.5 & 109.9 & 87.0 & 24.7 \\
5937.811 & 22.0 & 1.07 & $-$1.89 & RC02 & 101.9 & 62.1 & 120.4 & 81.2 & 55.0 & 60.8 & 102.9 & 65.8 & 60.4 & 70.6 & \ldots & \ldots & 95.5 & 31.7 \\
5941.752 & 22.0 & 1.05 & $-$1.52 & RC02 & \ldots & 92.5 & \ldots & 115.4 & 95.8 & 90.3 & \ldots & 100.4 & 81.0 & 103.8 & \ldots & \ldots & \ldots & 57.2 \\
5978.540 & 22.0 & 1.87 & $-$0.50 & IK01 & \ldots & \ldots & \ldots & \ldots & 80.1 & \ldots & 112.8 & 88.3 & 80.1 & \ldots & \ldots & \ldots & 103.6 & 54.4 \\
5999.680 & 22.0 & 2.17 & $-$0.73 & RC02 & 53.6 & 30.3 & 61.3 & 43.5 & 22.1 & 30.3 & 55.1 & 34.5 & 29.8 & 36.8 & \ldots & \ldots & \ldots & 14.9 \\
6091.174 & 22.0 & 2.27 & $-$0.42 & RC02 & 80.5 & 54.0 & 92.0 & 67.6 & 44.6 & 54.1 & 79.7 & 58.7 & 51.5 & 59.5 & 99.0 & 97.8 & 70.3 & 28.3 \\
6126.220 & 22.0 & 1.07 & $-$1.42 & IK01 & \ldots & 102.9 & \ldots & 124.5 & 106.6 & 102.1 & \ldots & 112.2 & 102.4 & 113.5 & \ldots & \ldots & \ldots & 67.8 \\
6146.220 & 22.0 & 1.87 & $-$1.47 & RC02 & \ldots & \ldots & \ldots & \ldots & \ldots & 22.8 & 46.5 & 26.5 & \ldots & \ldots & \ldots & 70.2 & \ldots & 8.7 \\
6186.150 & 22.0 & 2.17 & $-$1.15 & RC02 & \ldots & 21.0 & \ldots & 29.7 & 17.6 & 20.0 & \ldots & 20.2 & 16.6 & 25.3 & \ldots & \ldots & 26.0 & \ldots \\
6312.240 & 22.0 & 1.46 & $-$1.55 & IK01 & 88.9 & 54.0 & 112.0 & 70.6 & 44.4 & 50.7 & 87.9 & 58.4 & 50.4 & 62.7 & 118.1 & 115.8 & 94.3 & 18.5 \\
6497.690 & 22.0 & 1.44 & $-$1.93 & RC02 & 63.1 & 30.7 & 80.8 & 45.9 & \ldots & 31.2 & 64.5 & 33.9 & 29.5 & 40.7 & 95.0 & \ldots & 54.5 & 11.9 \\
6508.140 & 22.0 & 1.43 & $-$1.98 & RC02 & 57.5 & 26.6 & 82.3 & 41.8 & 19.0 & 23.1 & 61.1 & 29.2 & 23.9 & 34.5 & 107.1 & \ldots & 47.8 & 10.4 \\
6554.224 & 22.0 & 1.44 & $-$1.22 & RC02 & 118.5 & 77.6 & \ldots & 100.1 & \ldots & 77.4 & 118.6 & 83.5 & 74.7 & 88.8 & \ldots & \ldots & 112.9 & 46.2 \\
6716.680 & 22.0 & 2.49 & $-$1.04 & RC02 & 19.3 & \ldots & 29.2 & 15.3 & \ldots & 9.1 & 22.5 & 10.8 & \ldots & 11.6 & 39.4 & 38.1 & 14.1 & \ldots \\
6743.124 & 22.0 & 0.90 & $-$1.63 & RC02 & \ldots & 100.0 & \ldots & 123.8 & 96.0 & 97.0 & \ldots & \ldots & 96.3 & 110.4 & \ldots & \ldots & \ldots & 62.9 \\
6090.218 & 23.0 & 1.08 & $-$0.06 & RT03 & \ldots & 100.2 & \ldots & 120.1 & 102.2 & \ldots & \ldots & 106.0 & \ldots & 110.1 & \ldots & \ldots & \ldots & 66.6 \\
6216.362 & 23.0 & 0.28 & $-$1.29 & PN00 & \ldots & 122.8 & \ldots & \ldots & 110.2 & 119.7 & \ldots & \ldots & 117.4 & \ldots & \ldots & \ldots & \ldots & 75.1 \\
6251.818 & 23.0 & 0.29 & $-$1.34 & PN00 & \ldots & 109.9 & \ldots & \ldots & 98.8 & 108.7 & \ldots & 117.1 & 105.2 & 124.7 & \ldots & \ldots & \ldots & 61.1 \\
6504.160 & 23.0 & 1.18 & $-$1.23 & PN00 & 59.4 & 33.2 & 71.5 & 45.7 & 28.3 & 31.5 & 59.0 & 33.5 & 30.0 & 39.0 & 87.0 & 82.8 & 47.1 & 14.3 \\
5783.090 & 24.0 & 3.32 & $-$0.50 & CM05 & 47.1 & \ldots & 48.2 & 41.0 & 19.9 & 28.9 & 45.4 & 35.9 & 31.4 & 36.3 & 54.3 & 51.6 & 40.1 & 19.3 \\
5783.890 & 24.0 & 3.32 & $-$0.29 & CM05 & 73.8 & 55.8 & 75.4 & 67.2 & 46.5 & 53.7 & 72.4 & 58.9 & 56.6 & 57.1 & 78.0 & 77.5 & 63.1 & 34.8 \\
5787.960 & 24.0 & 3.32 & $-$0.08 & CM05 & 69.3 & 52.8 & 71.1 & 62.2 & 44.3 & 53.0 & 68.1 & 55.7 & 51.1 & 56.5 & \ldots & \ldots & 60.3 & \ldots \\
5537.742 & 25.0 & 2.19 & $-$2.02 & PN00 & 106.3 & 66.7 & \ldots & 80.9 & 60.0 & 64.9 & 105.6 & 72.5 & \ldots & 80.1 & \ldots & \ldots & 89.4 & 34.7 \\
6013.527 & 25.0 & 3.07 & $-$0.25 & PN00 & \ldots & 101.5 & \ldots & 117.7 & 93.5 & 100.2 & \ldots & 100.3 & 97.8 & 110.9 & \ldots & \ldots & 123.4 & 65.7 \\
6016.667 & 25.0 & 3.08 & $-$0.22 & PN00 & \ldots & 104.1 & \ldots & 116.0 & 103.2 & 102.7 & 123.9 & 107.6 & 102.1 & 110.5 & \ldots & \ldots & 116.1 & 73.9 \\
6021.800 & 25.0 & 3.07 & 0.03 & PN00 & 124.6 & 107.5 & \ldots & 117.8 & 114.6 & 106.3 & 123.5 & 109.5 & 103.0 & 113.9 & \ldots & \ldots & 121.9 & 80.4 \\
5530.791 & 27.0 & 1.71 & $-$2.06 & PN00 & 92.1 & 70.3 & 104.6 & 82.2 & 79.7 & 65.7 & 91.5 & 73.3 & 67.4 & 76.7 & 103.4 & 103.0 & 90.0 & 45.3 \\
6455.034 & 27.0 & 3.63 & $-$0.25 & PN00 & 27.8 & 21.8 & 32.1 & 26.7 & 26.0 & 22.5 & 30.4 & 21.9 & 21.4 & 24.3 & 30.5 & 27.8 & 22.9 & 12.0 \\
6632.447 & 27.0 & 2.28 & $-$2.00 & PN00 & 46.7 & 32.9 & 54.7 & 42.8 & 36.8 & 31.9 & 47.6 & 34.2 & 30.7 & 39.1 & 55.4 & 52.2 & 37.9 & 17.4 \\
5578.730 & 28.0 & 1.68 & $-$2.64 & KB95 & \ldots & 104.6 & \ldots & 117.0 & \ldots & 107.1 & \ldots & \ldots & 106.6 & \ldots & \ldots & \ldots & \ldots & 87.2 \\
6007.310 & 28.0 & 1.68 & $-$3.34 & RC02 & \ldots & \ldots & \ldots & \ldots & \ldots & 65.1 & \ldots & 67.0 & 63.8 & 70.1 & \ldots & \ldots & 81.5 & 45.3 \\
6086.282 & 28.0 & 4.26 & $-$0.52 & RC02 & 35.6 & 33.6 & 37.8 & 36.1 & 35.5 & 32.2 & 35.1 & 32.4 & 30.8 & 34.2 & \ldots & 34.1 & 32.7 & \ldots \\
6108.120 & 28.0 & 1.68 & $-$2.45 & KB95 & \ldots & 114.5 & \ldots & \ldots & \ldots & 115.2 & \ldots & 119.2 & 114.3 & 121.2 & \ldots & \ldots & \ldots & 98.3 \\
6175.370 & 28.0 & 4.09 & $-$0.53 & RC02 & 43.7 & 41.2 & \ldots & 44.2 & 44.5 & 41.3 & 46.6 & 42.4 & 40.2 & 44.2 & \ldots & \ldots & 36.3 & 30.7 \\
6186.711 & 28.0 & 4.10 & $-$0.97 & RC02 & \ldots & 20.0 & \ldots & 27.6 & 23.7 & 23.4 & 34.3 & 25.1 & 25.9 & 26.4 & \ldots & \ldots & 21.4 & 12.8 \\
6204.604 & 28.0 & 4.09 & $-$1.14 & RC02 & 20.5 & 20.1 & 30.7 & 23.2 & 21.6 & 21.2 & 24.9 & \ldots & 20.0 & 22.6 & \ldots & \ldots & 17.3 & 10.5 \\
6223.990 & 28.0 & 4.10 & $-$0.99 & IK01 & \ldots & 22.1 & \ldots & 23.9 & 22.2 & 21.7 & 28.1 & \ldots & 21.5 & 21.0 & \ldots & \ldots & 17.5 & 13.4 \\
6635.122 & 28.0 & 4.42 & $-$0.83 & RC02 & 16.5 & 16.0 & 12.8 & 19.2 & 17.6 & 18.8 & 19.3 & 14.8 & 16.0 & 20.0 & 25.3 & 21.7 & 11.4 & 8.2 \\
5782.127 & 29.0 & 1.64 & $-$1.72 & CS02 & synth & synth & synth & synth & synth & synth & synth & synth & synth & synth & synth & synth & synth & synth \\
4722.153 & 30.0 & 4.03 & $-$0.34 & KB95 & \ldots & synth & synth & synth & synth & synth & synth & synth & synth & synth & synth & synth & synth & synth \\
7070.100 & 38.0 & 1.85 & $-$0.09 & LUCK & 16.7 & 11.5 & 28.9 & 14.0 & 10.2 & \ldots & 15.5 & \ldots & 14.4 & 11.1 & \ldots & \ldots & 9.6 & \ldots \\
5570.440 & 42.0 & 1.34 & 0.15 & SS00 & 61.7 & 36.8 & 70.9 & 52.4 & 32.2 & \ldots & 61.2 & 39.3 & 32.5 & 44.8 & 81.4 & 79.5 & 46.4 & 9.9 \\
5274.240 & 58.1 & 1.04 & $-$0.32 & LUCK & 49.0 & 37.0 & 40.5 & 42.2 & 44.8 & \ldots & 43.9 & 38.8 & 36.7 & 41.0 & 49.7 & 50.3 & 30.0 & 20.5 \\
5472.300 & 58.1 & 1.24 & $-$0.18 & LUCK & 31.0 & 20.4 & 33.1 & 24.3 & 21.5 & \ldots & 27.8 & 21.4 & 22.3 & 22.4 & 44.2 & 50.0 & 18.2 & 9.2 \\
5322.760 & 59.1 & 0.48 & $-$0.07 & ND03 & 57.5 & 42.2 & 53.3 & 44.0 & \ldots & 34.1 & 52.2 & 40.9 & \ldots & 38.0 & 62.6 & 65.4 & 52.3 & 24.6 \\
4567.610 & 60.1 & 0.20 & $-$1.31 & DL03 & 43.7 & 26.8 & 40.7 & 33.6 & 28.1 & 24.3 & 33.5 & 29.6 & 25.3 & 28.7 & 51.3 & 56.2 & 41.4 & 17.9 \\
4706.540 & 60.1 & 0.00 & $-$0.71 & DL03 & \ldots & 77.5 & 109.0 & 94.2 & 100.1 & 76.2 & \ldots & 84.4 & 79.7 & 86.7 & 117.8 & 120.1 & 101.4 & 66.2 \\
4859.030 & 60.1 & 0.32 & $-$0.44 & DL03 & 95.3 & 64.0 & 100.5 & 83.6 & 78.3 & 62.7 & 85.7 & \ldots & 91.1 & 69.1 & 109.7 & 98.3 & \ldots & 108.9 \\
4902.040 & 60.1 & 0.06 & $-$1.34 & DL03 & 80.2 & \ldots & \ldots & \ldots & \ldots & \ldots & \ldots & 57.5 & \ldots & 62.8 & \ldots & \ldots & 75.0 & 36.9 \\
5092.790 & 60.1 & 0.38 & $-$0.61 & DL03 & \ldots & 58.5 & \ldots & \ldots & 64.3 & \ldots & \ldots & \ldots & \ldots & \ldots & \ldots & \ldots & 74.4 & 44.3 \\
5249.580 & 60.1 & 0.98 & 0.20 & DL03 & 83.8 & 65.9 & 84.8 & 75.9 & \ldots & \ldots & 81.5 & 72.4 & 62.5 & 69.5 & 91.9 & 88.7 & 80.4 & 47.3 \\
5306.460 & 60.1 & 0.86 & $-$0.97 & DL03 & \ldots & 16.5 & 36.6 & 22.7 & 17.2 & \ldots & 32.4 & 17.9 & 14.3 & 14.4 & \ldots & \ldots & 24.0 & 9.1 \\
5311.450 & 60.1 & 0.99 & $-$0.42 & DL03 & 63.3 & 36.7 & 64.7 & 47.8 & 40.0 & \ldots & 60.2 & 39.3 & 34.0 & 43.9 & \ldots & \ldots & 40.3 & 19.9 \\
5485.700 & 60.1 & 1.26 & $-$0.12 & DL03 & 43.1 & 29.2 & \ldots & 34.6 & 28.7 & \ldots & 44.8 & 32.4 & \ldots & 31.5 & \ldots & \ldots & 33.3 & 20.6 \\
5740.860 & 60.1 & 1.16 & $-$0.53 & DL03 & 28.5 & 19.9 & \ldots & \ldots & \ldots & \ldots & 27.1 & 20.5 & 17.6 & 22.6 & 35.4 & \ldots & 19.0 & 11.2 \\
5811.570 & 60.1 & 0.86 & $-$0.86 & DL03 & 28.9 & 18.2 & 29.4 & 23.5 & 16.8 & \ldots & 27.0 & 20.3 & 19.3 & 18.2 & 39.7 & 40.8 & 24.5 & 11.5 \\
4536.510 & 62.1 & 0.10 & $-$1.28 & LD06 & 53.7 & 36.5 & \ldots & 45.1 & \ldots & \ldots & 51.9 & 39.9 & 35.8 & 39.9 & \ldots & \ldots & \ldots & 27.9 \\
4577.690 & 62.1 & 0.25 & $-$0.65 & LD06 & synth & synth & synth & synth & synth & synth & synth & synth & synth & synth & synth & synth & synth & synth \\
4642.230 & 62.1 & 0.38 & $-$0.46 & LD06 & 69.0 & \ldots & \ldots & 60.2 & 62.1 & 52.1 & 63.4 & 57.0 & 48.8 & 55.8 & 72.4 & 77.3 & 72.4 & 42.3 \\
4676.900 & 62.1 & 0.04 & $-$0.87 & LD06 & 68.6 & 55.1 & 74.8 & 65.7 & 64.6 & 53.9 & 59.9 & 60.1 & 53.9 & 59.9 & 83.5 & 82.7 & 82.9 & \ldots \\
4316.050 & 64.1 & 0.66 & $-$0.45 & DL06 & synth & synth & synth & synth & synth & synth & synth & synth & synth & synth & synth & synth & synth & synth \\
4483.330 & 64.1 & 1.06 & $-$0.42 & DL06 & synth & synth & synth & synth & synth & synth & synth & synth & synth & synth & synth & synth & synth & synth \\
4093.155 & 72.1 & 0.05 & $-$1.15 & LH07 & synth & synth & synth & \ldots & synth & synth & synth & synth & \ldots & synth & synth & \ldots & synth & synth \\
5989.045 & 90.1 & 0.19 & $-$1.41 & NZ02 & synth & synth & synth & synth & synth & synth & synth & synth & synth & synth & synth & synth & synth & synth \\
\enddata

\tablerefs{CM05 = \citet{cohen05}; CS02 = \citep{cunha02}; DL03 = \citet{nd}; 
DL06 = \citet{gd}; 
IK01 = \citet{M4}; KB95 = \citet{kurucz95}; LD06 = \citet{sm}; 
LH07 = \citet{hf}; 
LUCK = R.\ E.\ Luck (priv.\ comm.); ND95 = \citet{norris95c}; 
NZ02 = \citet{nilsson02}; PN00 = \citet{prochaska00}; 
RC02 = \citet{ramirez02}; RT03 = \citet{bdp03}; SS00 = \citet{smith00}}

\end{deluxetable}

\clearpage
\pagestyle{plaintop}
\setlength{\voffset}{0mm}
\begin{deluxetable}{lcc} 
\tabletypesize{\footnotesize}
\tablecolumns{3} 
\tablewidth{0pc} 
\tablecaption{Adopted solar abundances.\label{tab:solar}}
\tablehead{ 
\colhead{Species} &
\colhead{$\log\epsilon$(X)} &
\colhead{Source} 
}
\startdata
Si & 7.55 & 1 \\
Ca & 6.31 & 2 \\
Sc & 3.05 & 2 \\
Ti & 4.90 & 2 \\
V & 4.00 & 2 \\
Cr & 5.64 & 2 \\
Mn & 5.39 & 2 \\
Fe & 7.50 & 1 \\
Co & 4.92 & 2 \\
Ni & 6.23 & 2 \\
Cu & 4.21 & 1 \\
Zn & 4.60 & 1 \\
Sr & 2.92 & 1 \\
Mo & 1.92 & 1 \\
Ce & 1.58 & 1 \\
Pr & 0.58 & 3 \\
Nd & 1.45 & 3 \\
Sm & 1.00 & 3 \\
Gd & 1.11 & 3 \\
Hf & 0.88 & 3 \\
Th & 0.06 & 3 \\
\enddata

\tablerefs{
1 = \citet{grevesse98}; 
2 = \citet{asplund05}; 
3 = \citet{grevesse07}
}
\end{deluxetable}

\begin{deluxetable}{lrrrrrrrrrrr} 
\tabletypesize{\tiny}
\tablecolumns{12} 
\tablewidth{0pc} 
\tablecaption{Abundance ratios [X/Fe] for Ca to Zn.\label{tab:abund1}}
\tablehead{ 
\colhead{Star} &
\colhead{[Si/Fe]} &
\colhead{[Ca/Fe]} &
\colhead{[Sc/Fe]} &
\colhead{[Ti/Fe]} &
\colhead{[V/Fe]} &
\colhead{[Cr/Fe]} &
\colhead{[Mn/Fe]} &
\colhead{[Co/Fe]} &
\colhead{[Ni/Fe]} &
\colhead{[Cu/Fe]} &
\colhead{[Zn/Fe]}
}
\startdata
M4 L1411 & 0.58 & 0.43 & 0.17 & 0.42 & 0.18 & 0.11 & $-$0.16 & 0.04 & 0.10 & $-$0.27 & \ldots \\
M4 L1501 & 0.64 & 0.45 & 0.21 & 0.35 & 0.18 & 0.07 & $-$0.23 & 0.01 & 0.15 & $-$0.27 & 0.24 \\
M4 L1514 & 0.51 & 0.34 & 0.00 & 0.44 & 0.15 & 0.06 & \ldots & 0.00 & 0.09 & $-$0.33 & 0.27 \\
M4 L2307 & 0.54 & 0.47 & 0.17 & 0.39 & 0.17 & 0.09 & $-$0.23 & 0.01 & 0.14 & $-$0.31 & 0.19 \\
M4 L2406 & 0.46 & 0.19 & 0.04 & 0.16 & 0.04 & $-$0.12 & $-$0.36 & $-$0.03 & 0.06 & $-$0.25 & 0.30 \\
M4 L2617 & 0.61 & 0.44 & 0.04 & 0.38 & 0.23 & 0.05 & $-$0.23 & 0.01 & 0.10 & $-$0.31 & 0.15 \\
M4 L3209 & 0.60 & 0.44 & 0.02 & 0.50 & 0.28 & 0.14 & $-$0.15 & 0.07 & 0.17 & $-$0.2 & 0.15 \\
M4 L3413 & 0.44 & 0.42 & 0.13 & 0.41 & 0.20 & 0.11 & $-$0.24 & 0.01 & 0.05 & $-$0.22 & 0.13 \\
M4 L3624 & 0.62 & 0.47 & 0.15 & 0.38 & 0.21 & 0.11 & $-$0.25 & 0.03 & 0.14 & $-$0.22 & 0.19 \\
M4 L4511 & 0.68 & 0.43 & 0.17 & 0.33 & 0.15 & 0.05 & $-$0.24 & 0.01 & 0.10 & $-$0.23 & 0.17 \\
M4 L4611 & 0.70 & 0.50 & 0.20 & 0.59 & 0.25 & 0.13 & \ldots & $-$0.13 & 0.17 & $-$0.46 & 0.29 \\
M4 L4613 & 0.52 & 0.42 & 0.10 & 0.55 & 0.25 & 0.16 & \ldots & $-$0.03 & 0.14 & $-$0.31 & 0.20 \\
M5 IV-81 & 0.32 & 0.37 & 0.08 & 0.40 & 0.11 & 0.08 & $-$0.24 & $-$0.08 & $-$0.02 & $-$0.52 & $-$0.07 \\
M5 IV-82 & 0.32 & 0.37 & 0.19 & 0.33 & 0.19 & 0.21 & $-$0.32 & $-$0.05 & $-$0.01 & $-$0.52 & $-$0.02 \\
\enddata

\end{deluxetable}

\begin{deluxetable}{lrrrrrrrrr} 
\tabletypesize{\scriptsize}
\tablecolumns{10} 
\tablewidth{0pc} 
\tablecaption{Abundance ratios [X/Fe] for Sr to Th.\label{tab:abund2}}
\tablehead{ 
\colhead{Star} &
\colhead{[Sr/Fe]} &
\colhead{[Mo/Fe]} &
\colhead{[Ce/Fe]} &
\colhead{[Pr/Fe]} &
\colhead{[Nd/Fe]} &
\colhead{[Sm/Fe]} &
\colhead{[Gd/Fe]} &
\colhead{[Hf/Fe]} &
\colhead{[Th/Fe]} 
}
\startdata
M4 L1411 & 0.67 & 0.18 & 0.61 & 0.62 & 0.64 & 0.47 & 0.42 & 0.43 & 0.49 \\
M4 L1501 & 0.74 & 0.13 & 0.52 & 0.56 & 0.43 & 0.47 & 0.36 & 0.39 & 0.60 \\
M4 L1514 & 0.87 & 0.16 & 0.36 & 0.33 & 0.45 & 0.32 & 0.14 & 0.37 & 0.33 \\
M4 L2307 & 0.67 & 0.20 & 0.51 & 0.45 & 0.51 & 0.52 & 0.27 & \ldots & 0.50 \\
M4 L2406 & 0.69 & $-$0.02 & 0.27 & \ldots & 0.11 & 0.27 & 0.08 & 0.30 & 0.36 \\
M4 L2617 & \ldots & \ldots & \ldots & 0.43 & 0.31 & 0.43 & 0.33 & 0.40 & 0.66 \\
M4 L3209 & 0.71 & 0.24 & 0.52 & 0.51 & 0.50 & 0.41 & 0.35 & 0.40 & 0.51 \\
M4 L3413 & \ldots & 0.17 & 0.52 & 0.50 & 0.50 & 0.44 & 0.35 & 0.33 & 0.59 \\
M4 L3624 & 0.91 & 0.13 & 0.56 & \ldots & 0.44 & 0.41 & 0.31 & \ldots & 0.70 \\
M4 L4511 & 0.61 & 0.13 & 0.53 & 0.41 & 0.42 & 0.43 & 0.33 & 0.37 & 0.53 \\
M4 L4611 & \ldots & 0.36 & 0.48 & 0.44 & 0.35 & 0.66 & 0.21 & 0.39 & 0.35 \\
M4 L4613 & \ldots & 0.31 & 0.65 & 0.57 & 0.58 & 0.54 & 0.33 & \ldots & 0.41 \\
M5 IV-81 & 0.51 & 0.02 & 0.14 & 0.40 & 0.32 & 0.41 & 0.59 & 0.28 & 0.64 \\
M5 IV-82 & \ldots & $-$0.17 & 0.26 & 0.37 & 0.31 & 0.50 & 0.45 & 0.23 & 0.59 \\
\enddata

\end{deluxetable}

\begin{deluxetable}{lrrrc} 
\tabletypesize{\footnotesize}
\tablecolumns{5} 
\tablewidth{0pc} 
\tablecaption{Abundance Dependences on Model Parameters for 
M4 L2307.\label{tab:parvar}}
\tablehead{ 
\colhead{Species} &
\colhead{\teff+50} &
\colhead{log $g$+0.2} &
\colhead{$\xi_t$+0.2} &
\colhead{Total\tablenotemark{a}} 
}
\startdata
{\rm [Si/Fe]} & 0.01 & $-$0.03 & 0.02 & 0.04 \\
{\rm [Ca/Fe]} & 0.10 & $-$0.09 & $-$0.04 & 0.14 \\
{\rm [Sc/Fe]} & 0.02 & 0.01 & $-$0.08 & 0.09 \\
{\rm [Ti/Fe]} & 0.12 & $-$0.08 & 0.01 & 0.14 \\
{\rm [V/Fe]} & 0.13 & $-$0.07 & $-$0.01 & 0.15 \\
{\rm [Cr/Fe]} & 0.09 & $-$0.09 & 0.01 & 0.13 \\
{\rm [Mn/Fe]} & 0.08 & $-$0.08 & $-$0.08 & 0.13 \\
{\rm [Fe/H]} & $-$0.01 & 0.04 & $-$0.03 & 0.05 \\
{\rm [Co/Fe]} & 0.05 & $-$0.04 & 0.01 & 0.06 \\
{\rm [Ni/Fe]} & 0.03 & $-$0.04 & 0.01 & 0.05 \\
{\rm [Cu/Fe]\tablenotemark{b}} & 0.06 & $-$0.04 & $-$0.05 & 0.09 \\
{\rm [Zn/Fe]\tablenotemark{b}} & $-$0.01 & $-$0.02 & $-$0.04 & 0.05 \\
{\rm [Sr/Fe]} & 0.10 & $-$0.09 & 0.05 & 0.15 \\
{\rm [Mo/Fe]} & 0.12 & $-$0.08 & 0.01 & 0.14 \\
{\rm [Ce/Fe]} & 0.04 & 0.01 & 0.03 & 0.05 \\
{\rm [Pr/Fe]} & 0.04 & $-$0.01 & 0.03 & 0.04 \\
{\rm [Nd/Fe]} & 0.05 & $-$0.01 & $-$0.02 & 0.05 \\
{\rm [Sm/Fe]\tablenotemark{b}} & 0.04 & $-$0.02 & $-$0.02 & 0.04 \\
{\rm [Gd/Fe]\tablenotemark{b}} & 0.03 & $-$0.01 & 0.03 & 0.05 \\
{\rm [Hf/Fe]\tablenotemark{c}} & 0.03 & $-$0.01 & 0.03 & 0.04 \\
{\rm [Th/Fe]\tablenotemark{b}} & 0.05 & 0.01 & 0.04 & 0.06 \\
\enddata

\tablenotetext{a}{The total value is the quadrature sum of the three 
individual abundance dependences}
\tablenotetext{b}{For elements whose abundances were derived via 
synthetic spectra, we assumed an equivalent width that produced the 
final abundance}
\tablenotetext{c}{Since no abundance was derived for Hf, we assumed 
an equivalent width that produced the mean cluster abundance}

\end{deluxetable}

\begin{deluxetable}{lc} 
\tabletypesize{\scriptsize}
\tablecolumns{2} 
\tablewidth{0pc} 
\tablecaption{Mg isotope ratios.\label{tab:mgiso}}
\tablehead{ 
\colhead{Star} &
\colhead{$^{24}$Mg:$^{25}$Mg:$^{26}$Mg} 
}
\startdata
M4 L1411 & 80:10:10 \\
M4 L1501 & \ldots \\
M4 L1514 & 80:10:10 \\
M4 L2307 & 72:14:14 \\
M4 L2406 & \ldots \\
M4 L2617 & 74:13:13 \\
M4 L3209 & 74:13:13 \\
M4 L3413 & 80:10:10 \\
M4 L3624 & 78:11:11 \\
M4 L4511 & 80:10:10 \\
M4 L4611 & 80:10:10 \\
M4 L4613 & 78:11:11 \\
M5 IV-81 & 80:10:10 \\
M5 IV-82 & 78:11:11 \\
\enddata

\end{deluxetable}

\clearpage

\clearpage

\clearpage

\clearpage

\clearpage

\clearpage

\clearpage

\clearpage

\clearpage

\clearpage

\clearpage

\clearpage

\clearpage

\end{document}